\pdfoutput=1 
\documentclass[twoside,fleqn,english,aps,pre,preprint,nofootinbib,a4paper]{revtex4-2}
\usepackage[T1]{fontenc}
\usepackage[latin9]{inputenc}
\usepackage{babel}
\usepackage{array}
\usepackage{units}
\usepackage{textcomp}
\usepackage{bm}
\usepackage{amsmath}
\usepackage{amssymb}
\usepackage{mathdots}
\usepackage{graphicx}
\usepackage{esint}
\usepackage[unicode=true,pdfusetitle,
 bookmarks=true,bookmarksnumbered=false,bookmarksopen=false,
 breaklinks=false,pdfborder={0 0 1},backref=false,colorlinks=false]
 {hyperref}

\makeatletter

\providecommand{\tabularnewline}{\\}

\numberwithin{equation}{section}

\usepackage{bm}
\usepackage{color}

\newcommand{\CC}[1]{{}}

\newcommand{\bS}{\begin{subequations}}
\newcommand{\eS}{\end{subequations}}

\DeclareMathOperator{\Pf}{Pf}
\DeclareMathOperator{\Tr}{tr}

\DeclareMathOperator{\arsinh}{arsinh}
\DeclareMathOperator{\arcosh}{arcosh}
\DeclareMathOperator{\artanh}{artanh}
\DeclareMathOperator{\arcoth}{arcoth}
\DeclareMathOperator{\csch}{csch}
\DeclareMathOperator{\gd}{gd}

\DeclareMathOperator{\Jam}{am}
\DeclareMathOperator{\Jss}{ss}\DeclareMathOperator{\Jcs}{cs}\DeclareMathOperator{\Jds}{ds}\DeclareMathOperator{\Jns}{ns}
\DeclareMathOperator{\Jsc}{sc}\DeclareMathOperator{\Jcc}{cc}\DeclareMathOperator{\Jdc}{dc}\DeclareMathOperator{\Jnc}{nc}
\DeclareMathOperator{\Jsd}{sd}\DeclareMathOperator{\Jcd}{cd}\DeclareMathOperator{\Jdd}{dd}\DeclareMathOperator{\Jnd}{nd}
\DeclareMathOperator{\Jsn}{sn}\DeclareMathOperator{\Jcn}{cn}\DeclareMathOperator{\Jdn}{dn}\DeclareMathOperator{\Jnn}{nn}

\newcommand{\cI}[1]{(I.#1)}

\makeatother

\begin{document}
\global\long\def\mat#1{\bm{\mathrm{#1}}}%

\CC{
\global\long\def\Pf{\,\mathrm{Pf}\,}%
\global\long\def\Tr{\mathrm{tr}}%

\global\long\def\arsinh{\mathrm{arsinh}}%
\global\long\def\arcosh{\mathrm{arcosh}}%
\global\long\def\artanh{\mathrm{artanh}}%
\global\long\def\arcoth{\mathrm{arcoth}}%
\global\long\def\csch{\mathrm{csch}}%
\global\long\def\arccot{\mathrm{arccot}}%
\global\long\def\gd{\mathrm{gd}}%

\global\long\def\Jam{\operatorname{am}}%

\global\long\def\Jss{\operatorname{ss}}%
\global\long\def\Jcs{\operatorname{cs}}%
\global\long\def\Jds{\operatorname{ds}}%
\global\long\def\Jns{\operatorname{ns}}%

\global\long\def\Jsc{\operatorname{sc}}%
\global\long\def\Jcc{\operatorname{cc}}%
\global\long\def\Jdc{\operatorname{dc}}%
\global\long\def\Jnc{\operatorname{nc}}%

\global\long\def\Jsd{\operatorname{sd}}%
\global\long\def\Jcd{\operatorname{cd}}%
\global\long\def\Jdd{\operatorname{dd}}%
\global\long\def\Jnd{\operatorname{nd}}%

\global\long\def\Jsn{\operatorname{sn}}%
\global\long\def\Jcn{\operatorname{cn}}%
\global\long\def\Jdn{\operatorname{dn}}%
\global\long\def\Jnn{\operatorname{nn}}%

see preamble }

\global\long\def\Def{\equiv}%
\let\Def=\equiv

\global\long\def\T#1{#1^{\top}}%
\global\long\def\CT#1{#1^{\dagger}}%
\global\long\def\Tc{T_{\mathrm{c}}}%
\global\long\def\K{\mathcal{K}}%
\global\long\def\P#1{P_{\boldsymbol{#1}}}%

\global\long\def\ii{\mathrm{i}}%
\global\long\def\ee{\mathrm{e}}%
\global\long\def\dd{\mathrm{d}}%

\global\long\def\L{\leftrightarrow}%
\global\long\def\M{\updownarrow}%
\global\long\def\xM{x_{\M}}%
\global\long\def\ui{\check{u}}%

\global\long\def\fliptrans{\mathcal{S}}%
\global\long\def\lan{{\lambda_{\mathrm{n}}}}%
\global\long\def\las{{\lambda_{\mathrm{s}}}}%
\global\long\def\lac{{\lambda_{\mathrm{c}}}}%
\global\long\def\lad{{\lambda_{\mathrm{d}}}}%

\renewcommand*{\arraystretch}{0.8}
\title{The square lattice Ising model on the rectangle III: \\
Hankel and Toeplitz determinants}
\author{Alfred Hucht}
\affiliation{Faculty of Physics, University of Duisburg-Essen, 47048 Duisburg,
Germany}
\date{\today}
\begin{abstract}
Based on the results obtained in {[}\href{https://doi.org/10.1088/1751-8121/aa5535}{Hucht, J. Phys. A: Math. Theor. 50, 065201 (2017)}{]},
we show that the partition function of the anisotropic square lattice
Ising model on the $L\times M$ rectangle, with open boundary conditions
in both directions, is given by the determinant of a $\nicefrac{M}{2}\times\nicefrac{M}{2}$
Hankel matrix, that equivalently can be written as the Pfaffian of
a skew-symmetric $M\times M$ Toeplitz matrix. The $M-1$ independent
matrix elements of both matrices are Fourier coefficients of a certain
symbol function, which is given by the ratio of two characteristic
polynomials. These polynomials are associated to the different directions
of the system, encode the respective boundary conditions, and are
directly related through the symmetry of the considered Ising model
under exchange of the two directions. The results can be generalized
to other boundary conditions and are well suited for the analysis
of finite-size scaling functions in the critical scaling limit using
Szeg\H{o}'s theorem.
\end{abstract}
\maketitle
\newpage{}

\tableofcontents{}

\section{Introduction}

\begin{samepage}

The anisotropic two-dimensional Ising model \citep{Ising25} on the
$L\times M$ square lattice is one of the best investigated models
in statistical mechanics. In the thermodynamic limit $L,M\to\infty$,
it has a continuous phase transition, from a disordered high-temperature
phase to an ordered low-temperature phase, at a critical temperature
$\Tc$. After the exact solution of the periodic case by Onsager \citep{Onsager44,Kaufman49},
many authors have contributed to the knowledge about this model under
various aspects, such as different boundary conditions (BCs) or surface
effects \citep{McCoyWu73,Baxter82,Abraham86}.

\end{samepage}

Until some years ago, exact solutions for arbitrary temperatures were
only known for systems with periodic or antiperiodic boundary conditions
in at least one direction, as then a Fourier transform along this
translationally invariant direction could be used to diagonalize the
problem in the corresponding direction. The remaining direction could
be handled afterwards with a transfer matrix method, involving a $2\times2$
transfer matrix, taking into account arbitrary boundary conditions,
or even line disorder \citep{McCoyWu73}. 

This changed in 2016, when Baxter \citep{Baxter16,Baxter20} and Hucht
\citep{Hucht16a,Hucht16ae,Hucht16b} independently presented exact
results for the Ising model on the rectangle, with open boundary conditions
in both directions, expressing the partition function as $M\times M$
and $M/2\times M/2$ determinants, respectively. While Baxter used
Kaufman's spinor method \citep{Kaufman49} below $\Tc$ and focused
on the thermodynamic limit \citep{Baxter16}, where he exactly calculated
the corner contributions to the free energy conjectured in \citep{VernierJacobsen12},
Hucht utilized the dimer method of Kasteleyn and McCoy \& Wu \citep{Kasteleyn61,Kasteleyn63,Fisher66,McCoyWu73,McCoyWu14},
combined with Schur reductions and a block transfer matrix formulation
of Molinari \citep{Molinari08}, and derived closed expressions for
finite systems at arbitrary temperatures. The resulting transfer matrices
directly correspond to the spinor method matrices, bridging the gap
between these so far rather unrelated methods. 

Near the critical temperature $\Tc$, the direction-dependent bulk
correlation length of thermal fluctuations $\xi_{\infty}^{\delta}(T)$
diverges according to $\xi_{\infty}^{\delta}(T\,{>}\,\Tc)\simeq\xi_{0}^{\delta}(T/\Tc-1)^{-\nu}$,
where $\delta=\,\L,\M$ denotes the directions corresponding to $L$
and $M$, $\xi_{0}^{\delta}$ are interaction-dependent metric factors,
and the correlation length critical exponent is $\nu=1$ in the two-dimensional
Ising model\footnote{The relation ``$\simeq$'' denoted ``asymptotically equal'' in
the respective limit, and $\Def$ denotes a definition.}. If $\xi_{\infty}^{\L,\M}(T)$ is of the order of (or larger as)
the respective system size $L$ or $M$, interesting finite-size effects
such as the critical Casimir effect emerge, which describes interactions
of the system boundaries mediated by long-range critical fluctuations
\citep{FisherdeGennes78,FisherAu-Yang80} in close analogy to the
quantum electrodynamical Casimir effect \citep{CasimirPolder48,Casimir48}.
These finite-size effects can be described \citep{HuchtGruenebergSchmidt11}
by universal finite-size scaling functions of the form $\Theta_{\M}(\xM,\rho)$,
with temperature scaling variable $\xM\Def(T/\Tc-1)(M/\xi_{0}^{\M})^{1/\nu}$
and generalized aspect ratio $\rho\Def(L/\xi_{0}^{\L})/(M/\xi_{0}^{\M})$,
that only depend on the bulk and surface universality classes of the
model \citep{Kadanoff66,Diehl97a}, as well as on the BCs. They have
been calculated exactly for many cases, albeit mostly in strip geometry,
where the aspect ratio $\rho$ of the system goes to zero \citep{EvansStecki94,Au-YangFisher80,BrankovDantchevTonchev00,Gambassi09a,RudnickZandiShackellAbraham10,AbrahamMaciolek10,AbrahamMaciolek13}.
The theoretical results from exact calculations, field theory and
Monte Carlo simulations \citep{AbrahamMaciolek10,AbrahamMaciolek13,Gambassi09a,Hasenbusch0905,Hasenbusch0907,Hasenbusch0908,Hasenbusch1005,Hasenbusch1104,Hasenbusch1205,Hucht07a,MaciolekGambassiDietrich07,RudnickZandiShackellAbraham10,VasilyevGambassiMaciolekDietrich07,VasilyevGambassiMaciolekDietrich09,HobrechtHucht14}
were shown to be in excellent agreement with experiments \citep{GarciaChan99url,GarciaChan00a,GarciaChan00b,GarciaChan02,FukutoYanoPershan05,GanshinScheidemantelGarciaChan06,HertleinHeldenGambassiDietrichBechinger08,GamMacHerNelHelBechDiet09}
for various bulk and surface universality classes.

\begin{figure}
\begin{centering}
\includegraphics[width=0.65\textwidth]{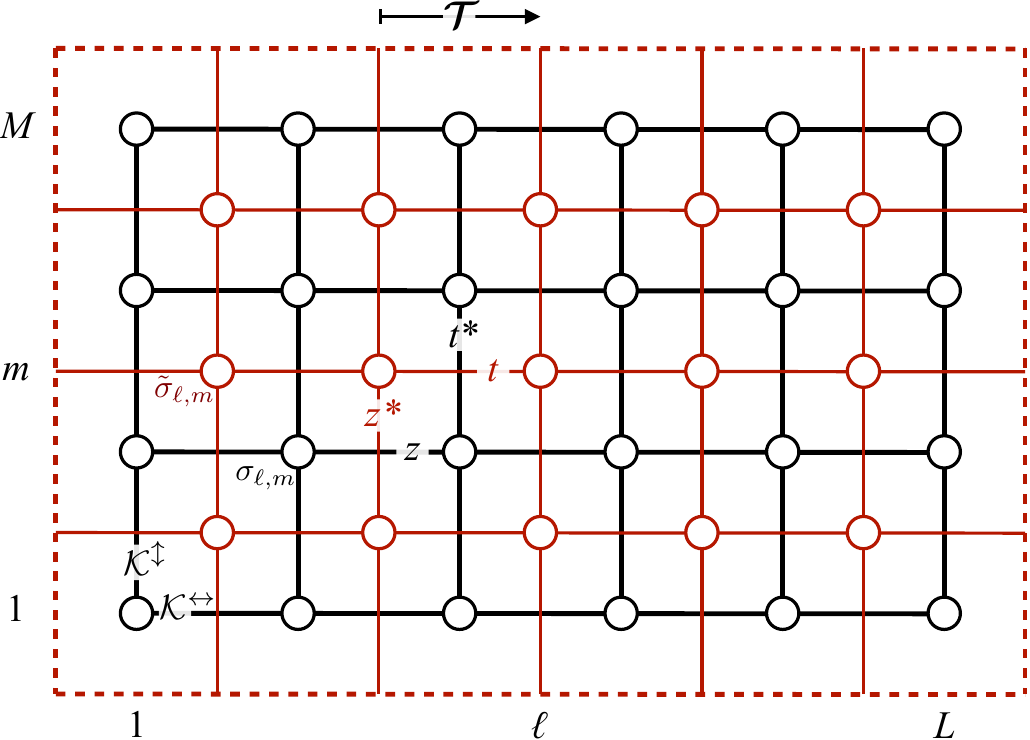}
\par\end{centering}
\caption{\linespread{1.2}\selectfont{}The anisotropic square lattice Ising
model on the $L\times M$ rectangle (black), together with its dual
lattice (red) and with the relevant couplings (see text). The transfer
matrix $\protect\mat{\mathcal{T}}$ (\ref{eq:T_2}) propagates as
indicated. \label{fig:lattice}}
\end{figure}
Directly at the critical point, exact methods or conformal field theory
\citep{Polyakov70,Cardy84,BurkhardtEisenriegler95,Cardy2006,BimonteEmigKardar13}
can be used to get exact expressions for the Casimir amplitude $\Delta_{\mathrm{C}}(\rho)$
for arbitrary $\rho$. This has been done for periodic \citep{FerdinandFisher69,LuWu01}
as well as for open BCs \citep{KlebanVassileva91,WuIzmailianGuo12}.
Using conformal maps, these results can be used to investigate fluctuation-induced
forces between colloids in critical suspensions both in theory \citep{SchlesenerHankeDietrich03,KHD09,TrKoGamHarDiet09,GambassiDietrich10,TrZvGamVogtHarBechDiet11,Hasenbusch13,LTHD14,HobrechtHucht15a,ETBEvRD15}
and experiment \citep{BrunnerDobnikar04,SoykaZvyaHertHeldBech08,BonnOtwiSacaGuoWegSchall09,ZAB11,GZS12,GZTS12,DVNBS13,NguyenFaHuWeSch13,TasiosDijkstra17}.
At arbitrary aspect ratios and temperatures, however, the finite-size
scaling functions must be derived from the exact solution of the finite
system with the correct BCs. For the Ising model, this has been done
only in a few cases, namely for the torus with periodic BCs in both
directions \citep{HuchtGruenebergSchmidt11,HobrechtHucht16a,HobrechtHucht18a},
for the cylinder with open or fixed at the boundaries in one direction
\citep{HobrechtHucht16a,HobrechtHucht18b}, and for the open rectangle
considered here \citep{Hucht16b}. 

In this work, the calculations of \citep{Hucht16a} are substantially
simplified by (i) the use of elliptic functions \citep{Baxter82},
(ii) a simplified normalization of the eigenvectors of the relevant
$M\times M$ transfer matrices, (iii) a transformation to Hankel and
Toeplitz matrices, and (iv) a representation of the sums over transfer
matrix eigenvalues through a complex contour integral, having (v)
a remarkably simple integrand given by the ratio of two very simple
characteristic polynomials, each representing one direction of the
two-dimensional Ising model. The preceding publication \citep{Hucht16a}
will be denoted I in the following. 

\subsection{The model}

This work focuses on the anisotropic square lattice Ising model on
the $L\times M$ open rectangle, shown in figure~\ref{fig:lattice}.
Our aim is to calculate the partition function 
\begin{equation}
Z=\Tr\exp\left(\K^{\L}\sum_{\ell=1}^{L-1}\sum_{m=1}^{M}\sigma_{\ell,m}\sigma_{\ell+1,m}+\K^{\M}\sum_{\ell=1}^{L}\sum_{m=1}^{M-1}\sigma_{\ell,m}\sigma_{\ell,m+1}\right),\label{eq:Ising}
\end{equation}
with reduced couplings $\K^{\delta}=\beta J^{\delta}$ in direction
$\delta=\,\L,\M$, where the trace is over all $2^{LM}$ configurations
of the $LM$ spins $\sigma_{\ell,m}=\pm1$. We assume open BCs both
in horizontal ($L$) and in vertical ($M$) directions, and we assume
even $M$. The starting point of this work is the matrix product representation
\cI{29} of the well known Pfaffian representation by Kasteleyn, McCoy
\& Wu \citep{Kasteleyn63,McCoyWu73}, derived in \citep{Hucht16a}
using a Schur reduction as well as a block transfer matrix formula
by Molinari \citep{Molinari08}. We follow the notation of \citep{Hucht16a},
with a few clarifying minor modifications. Therefore, we define the
dual 
\begin{equation}
a^{*}\Def\frac{1-a}{1+a}\label{eq:dual}
\end{equation}
of some quantity $a$, as well as the quite useful abbreviations\footnote{Subscripts $a_{\pm}$ are consequently used in this manner, while
superscripts $a^{\pm}$ may have different meanings.}
\begin{equation}
a_{\pm}\Def\frac{a\pm a^{-1}}{2}\label{eq:apm}
\end{equation}
introduced in \cI{20}, fulfilling $(a^{*})^{*}=a=a_{+}+a_{-}$. The
reduced couplings $\K^{\delta}$ are rewritten using the weights $z$
and $t$ according to
\begin{equation}
z\Def\tanh\K^{\L},\qquad t\Def(\tanh\K^{\M})^{*}=\ee^{-2\K^{\M}}.\label{eq:z_t_Def}
\end{equation}
The partition function $Z$ of the considered anisotropic Ising model
with open BCs in both directions is invariant under exchange of the
two directions $\L$ and $\M$, where both the system dimensions and
the coupling constants are exchanged according to the \emph{swap transformation}%
\begin{equation}
\fliptrans:(L,M;\K^{\L},\K^{\M})\mapsto\fliptrans[(L,M;\K^{\L},\K^{\M})]=(M,L;\K^{\M},\K^{\L}),\label{eq:flip_trans}
\end{equation}
such that $Z=\fliptrans[Z]$. The swap transformation is a involution,
as $\fliptrans[\fliptrans[a]]=a$. We now summarize the required results
from reference~\citep{Hucht16a}. 

\subsection{Starting point}

In equations~(27-29) of \citep{Hucht16a} we showed that the square
of the partition function (\ref{eq:Ising}) is proportional to the
determinant of a matrix product of the form $\langle\mat e_{2}|\mat{\mathcal{T}}_{2}^{L}|\mat e_{2}\rangle$,
with real hyperbolic $2\times2$ block transfer matrix $\mat{\mathcal{T}}_{\!2}$
propagating in horizontal direction. In this work, we will use a ``Wick
rotated'' version of $\mat{\mathcal{T}}_{\!2}$ according to\footnote{$\T{\mat A\negmedspace}$ and $\CT{\mat A}$ denote the transpose
and the conjugate transpose of $\mat A$, respectively.} $\mat{\mathcal{T}}\Def\mat{\mathcal{I}}\,\mat{\mathcal{T}}_{\!2}\,\CT{\mat{\mathcal{I}}},$
with matrix $\mat{\mathcal{I}}\Def(\begin{smallmatrix}1 & 0\\
0 & \ii
\end{smallmatrix})\otimes\mat 1$, leading to the complex orthogonal block transfer matrix 
\begin{equation}
\mat{\mathcal{T}}\Def\begin{bmatrix}\mat T_{+} & -\ii\mat T_{-}\\
\,\ii\mat T_{-} & \mat T_{+}
\end{bmatrix},\label{eq:T_2}
\end{equation}
which fulfills $\mat{\mathcal{T}}^{-1}=\T{\mat{\mathcal{T}}}$. Note
that $\mat{\mathcal{T}}$ is not unitary. The two real symmetric $M\times M$
matrices $\mat T_{\pm}$ were given in \cI{30} and can be simplified
to \bS\begingroup\allowdisplaybreaks\label{eqs:TpTm}
\begin{align}
\mat T_{+} & =-\frac{t_{-}z_{-}}{2}{\setlength{\arraycolsep}{6pt}\,\,\underbrace{\!\!\begin{pmatrix}2{+}\frac{t^{*}}{z^{*}} & 1\\
1 & 2 & \ddots\\
 & \ddots & \ddots & \ddots\\
 &  & \ddots & 2 & 1\\
 &  &  & 1 & 2{+}t^{*}z^{*}
\end{pmatrix}\!\!}_{{\textstyle \mat C}}}\,\,+(t_{+}z_{+}+t_{-}z_{-})\mat 1,\label{eq:Tp}\\
\mat T_{-} & =-\frac{t_{-}z_{-}}{2}{\setlength{\arraycolsep}{5pt}\begin{pmatrix} &  &  & z^{*} & -\frac{1}{t^{*}}\\
 &  & \iddots & -2{t^{*}}_{\!+} & \frac{1}{z^{*}}\\
 & \iddots & \iddots & \iddots\\
z^{*} & -2{t^{*}}_{\!+} & \iddots\\
-\frac{1}{t^{*}} & \frac{1}{z^{*}}
\end{pmatrix}},\label{eq:Tm}
\end{align}
\endgroup\eS{}using the dual couplings $z^{*}$ and $t^{*}$ (the
matrix $\mat C$ will be utilized later). They are related via the
transfer matrix $\mat T=\mat T_{+}+\mat T_{-}$ according to (\ref{eq:apm}),
\begin{equation}
\mat T_{\pm}=\tfrac{1}{2}\left(\mat T\pm\mat T^{-1}\right)\quad\Leftrightarrow\quad\mat T^{\pm1}=\mat T_{+}\pm\mat T_{-},\label{eq:T_pm}
\end{equation}
from which directly follows that $\mat{\mathcal{T}}$ can be block
diagonalized through a rotation by $\pi/4$,
\begin{equation}
\mat{\mathcal{R}}_{\frac{\pi}{4}}\CT{\mat{\mathcal{I}}}\mat{\mathcal{T}}\,\mat{\mathcal{I}}\,\mat{\mathcal{R}}_{-\frac{\pi}{4}}=\begin{bmatrix}\mat T & \mat 0\\
\mat 0 & \mat T^{-1}
\end{bmatrix},\label{eq:block_diagonal}
\end{equation}
with block rotation matrix $\mat{\mathcal{R}}_{\theta}\Def(\begin{smallmatrix}\cos\theta & \sin\theta\\
-\sin\theta & \cos\theta
\end{smallmatrix})\otimes\mat 1$, for details see chapter V in \citep{Hucht16a}. The open BCs in
horizontal direction are represented through the boundary state
\begin{equation}
|\mat e_{\mathrm{o}}\rangle\Def\frac{1}{\sqrt{2}}|\begin{array}{cc}
\mat 1 & \ii\mat S\end{array}\rangle,\qquad\langle\mat e_{\mathrm{o}}|\mat e_{\mathrm{o}}\rangle=\mat 1,\label{eq:vec_e}
\end{equation}
where $|\mat e_{\mathrm{o}}\rangle$ is a two element block vector,
with the $M\times M$ matrices 
\begin{align}
\mat 1\Def & \left(\begin{array}{ccc}
1\\
 & \;\;\ddots\;\;\\
 &  & 1
\end{array}\right), & \mat S\Def & \left(\begin{array}{ccc}
 &  & 1\\
 & \;\;\iddots\;\;\\
1
\end{array}\right).\label{eq:1_and_S}
\end{align}
Together with the constant%
\begin{equation}
Z_{0}\Def(1-z^{2})^{\frac{M}{2}}\left(\frac{2}{-z_{-}}\right)^{\negmedspace\frac{LM}{2}}\!,\label{eq:Z0}
\end{equation}
the square of the partition function (\ref{eq:Ising}) then reads
\begin{equation}
Z^{2}=Z_{0}^{2}\det\langle\mat e_{\mathrm{o}}|\mat{\mathcal{T}}^{L}|\mat e_{\mathrm{o}}\rangle,\label{eq:det(T2)}
\end{equation}
see also \citep[(2.35)]{Baxter16}. Defining the projectors 
\begin{equation}
\mat S^{\pm}\Def\frac{1}{2}\left(\mat 1\pm\mat S\right),\label{eq:Spm}
\end{equation}
the argument of the determinant in (\ref{eq:det(T2)}) becomes \bS\label{eq:MM}
\begin{align}
\langle\mat e_{\mathrm{o}}|\mat{\mathcal{T}}^{L}|\mat e_{\mathrm{o}}\rangle & =\langle\begin{array}{cc}
\mat S^{+} & \ii\mat S^{-}\end{array}|\begin{bmatrix}\mat T{}^{L} & \mat 0\\
\mat 0 & \mat T{}^{-L}
\end{bmatrix}|\begin{array}{cc}
\mat S^{+} & \ii\mat S^{-}\end{array}\rangle\label{eq:MM_1}\\
 & =\mat S^{+}\,\mat T{}^{L}\,\mat S^{+}+\mat S^{-}\,\mat T{}^{-L}\,\mat S^{-}=\mat{\T M}\mat M,\label{eq:MM_2}
\end{align}
\eS{}with the matrix
\begin{equation}
\mat M\Def\mat x\big(\mat T^{L/2}\,\mat S^{+}+\mat T^{-L/2}\,\mat S^{-}\big),\label{eq:M_def}
\end{equation}
as $\mat{\mat S^{+}\mat S^{-}=\mat 0}$. In the following we will
determine the eigenvalues $\lambda_{\mu},\lambda_{\pm,\mu}$ and common
eigenvectors $\vec{x}_{\mu}=(\mat x)_{\mu}$ of $\mat T$ and $\mat T_{\pm}$,
which fulfill
\begin{align}
\mat T\,\vec{x}_{\mu} & =\lambda_{\mu}\vec{x}_{\mu}, & \mat T_{\pm}\vec{x}_{\mu} & =\lambda_{\pm,\mu}\vec{x}_{\mu},\label{eq:eigensystem}
\end{align}
where $\mu=1,\ldots,M$.

\subsection{The Onsager dispersion relation}

\label{subsec:Onsager-dispersion-relation}

The next step in the calculation of the partition function (\ref{eq:Ising})
is the determination of the eigenvalues $\lambda_{+,\mu}$ as zeroes
of the characteristic polynomial (CP) of the tridiagonal matrix $\mat T_{+}$
from (\ref{eq:Tp}), which will lead to the Onsager dispersion relation.
As we will discuss several characteristic polynomials (CPs) in the
following, we first define the CP $\P a(x)$ of an arbitrary $M\times M$
matrix $\mat A$, with eigenvalues $(\boldsymbol{a})_{\mu}=a_{\mu}\in\mathbb{C}$,
to be
\begin{equation}
\P a(x)\Def\det(x\mat 1-\mat A)=\prod_{\mu=1}^{M}(x-a_{\mu}),\label{eq:CP_Def}
\end{equation}
such that $\P a(x)$ is a polynomial of degree $M$ in the indeterminate
$x\in\mathbb{C}$. 

Using the well known recursion formula for tridiagonal matrices (see,
e.g., \citep{Molinari08}), in \citep{Hucht16a} we derived\footnote{The sign change in the definition w.r.t.~\cI{9} has no effect for
even $M$.}\bS\label{eqs:CP+}
\begin{align}
\P{\lambda_{+}}(\lambda_{+}) & =\det(\lambda_{+}\mat 1-\mat T_{+})=\prod_{\mu=1}^{M}(\lambda_{+}-\lambda_{+,\mu})\label{eq:CP+_Def}\\
 & =\left(\frac{t_{-}z_{-}}{2}\right)^{\negmedspace M}\langle\begin{array}{cc}
1/z^{*} & -t^{*}\end{array}|\,\mat Q^{M}\,|\begin{array}{cc}
z^{*} & t^{*}\end{array}\rangle,\label{eq:CP+_2}
\end{align}
\eS{}where $\lambda_{+}\in\mathbb{C}$. We point out the obvious
similarity between equations (\ref{eq:det(T2)}) and (\ref{eq:CP+_2}),
which will become clearer later. The vertically propagating $2\times2$
transfer matrix
\begin{equation}
\mat Q\Def\begin{pmatrix}2\frac{t_{+}z_{+}-\lambda_{+}}{t_{-}z_{-}} & -1\\
1 & 0
\end{pmatrix}=\begin{pmatrix}2\zeta_{+} & -1\\
1 & 0
\end{pmatrix}=\begin{pmatrix}2\cos\varphi\, & -1\\
1 & 0
\end{pmatrix}\label{eq:Q}
\end{equation}
has the eigenvalues $\zeta^{\pm1}$ with modulus one\footnote{$\zeta^{\pm1}$ were denoted $q^{\pm}$ in \cI{42}.},
such that the $n$-th power of $\mat Q$ reads 
\begin{equation}
\mat Q^{n}=\frac{1}{\sin\varphi}\begin{pmatrix}\sin([n+1]\varphi) & -\sin(n\varphi)\\
\sin(n\varphi) & -\sin([n-1]\varphi)
\end{pmatrix}.\label{eq:Qn}
\end{equation}
Here and in the following, we express the horizontal and vertical
eigenvalues $(\lambda,\zeta)$ through the introduced angles $(\gamma,\varphi)$
according to \bS\label{eqs:eigenvalues}
\begin{align}
\lambda & =\ee^{\gamma}, & \lambda_{+} & =\cosh\gamma, & \lambda_{-} & =\sinh\gamma,\label{eq:gamma}\\
\zeta & =\ee^{\ii\varphi}, & \zeta_{+} & =\cos\varphi, & \zeta_{-} & =\ii\sin\varphi.\label{eq:zeta}
\end{align}
\eS{}Equation (\ref{eq:Q}) is the point in the calculation where
the famous Onsager dispersion relation%
\begin{equation}
\lambda_{+}+t_{-}z_{-}\zeta_{+}=t_{+}z_{+}\qquad\Leftrightarrow\qquad\cosh\gamma+t_{-}z_{-}\cos\varphi=t_{+}z_{+}\label{eq:ODR}
\end{equation}
between the eigenvalues $\lambda$ and $\zeta$ enters the stage,
which plays the key role in the exact solution of the square lattice
Ising model \citep{Onsager44}. It relates the two ``good'' variables
$\lambda$ and $\zeta$ for propagation in $\L$ ($L$) and $\M$
($M$) direction, respectively. As pointed out by Baxter \citep[chap. 15.10]{Baxter82},
a parametrization of relation (\ref{eq:ODR}) using elliptic functions
considerably simplifies the analysis. This parametrization is introduced
in the next chapter, and we return to the characteristic polynomials
in chapter \ref{sec:CharPoly}.

\section{Elliptic parametrization}

The key idea behind the elliptic parametrization of the Onsager dispersion
relation (\ref{eq:ODR}) is (i) to substitute the coupling constants
$(z,t)$ through new constants $(k,\eta)$, where $k$ is temperature-like
and $\eta$ encodes the coupling anisotropy, and (ii) to introduce
a complex variable $u$ that simultaneously parametrizes the quantities
$\zeta=\zeta(u)$ and $\lambda=\lambda(u)$ in such a way that (\ref{eq:ODR})
is always fulfilled\footnote{Imagine the simpler case of a circle $x^{2}+y^{2}=r^{2}$ being trigonometrically
parametrized, with $x(u)=r\cos u$, $y(u)=r\sin u$ and complex parameter
$u$.}. We recapitulate the parametrization of (\ref{eq:ODR}) through the
Jacobi elliptic functions $\Jsn(u,k)$, $\Jcn(u,k)$ and $\Jdn(u,k)$
\citep{Lawden89,NIST,NIST:DLMF}, with elliptic modulus $k$, by first
defining the Jacobi amplitude 
\begin{equation}
\phi\Def\Jam(u,k)\label{eq:am_Def}
\end{equation}
as the inverse function of the elliptic integral of the first kind
\begin{equation}
u=F(\phi,k)\Def\int_{0}^{\phi}(1-k^{2}\sin^{2}\theta)^{-1/2}\dd\theta,\label{eq:F(phi,k)}
\end{equation}
for $\{\phi,u,k\}\in\mathbb{R}.$ Consequently, the Jacobi elliptic
functions are given by 
\begin{equation}
\Jsn(u,k)\Def\sin\phi,\qquad\Jcn(u,k)\Def\cos\phi,\qquad\Jdn(u,k)\Def\frac{\partial\phi}{\partial u}=\pm\sqrt{1-k^{2}\sin^{2}\phi},\label{eq:sncndn}
\end{equation}
and fulfill the sum of squares identities
\begin{equation}
\Jsn^{2}(u,k)+\Jcn^{2}(u,k)=k^{2}\Jsn^{2}(u,k)+\Jdn^{2}(u,k)=1.\label{eq:sncndn_id}
\end{equation}
As common, we suppress the modulus $k$ if possible and write, e.\,g.,
$\Jsn u$ instead of $\Jsn(u,k)$. We follow Glashier's notation \citep[(22.2.10)]{NIST:DLMF}
and define all sixteen Jacobi elliptic functions 
\begin{equation}
\mathrm{pq}\,u\Def\frac{\mathrm{pr}\,u}{\mathrm{qr}\,u}=\frac{1}{\mathrm{qp}\,u},\qquad\mathrm{where}\qquad\mathrm{p},\mathrm{q},\mathrm{r}\in\{\mathrm{s},\mathrm{c},\mathrm{d},\mathrm{n}\},\label{eq:Glashiers_16}
\end{equation}
including the four trivial ones, $\Jss u=\Jcc u=\Jdd u=\Jnn u=1$. 

The Jacobi elliptic functions are double periodic and meromorphic
in the complex $u$-plane, that is they are analytic up to simple
poles. The common quarter-periodicity rectangle has the corners 
\begin{equation}
\{u_{\mathrm{s}},u_{\mathrm{c}},u_{\mathrm{d}},u_{\mathrm{n}}\}\Def\{0,K,K+\ii K',\ii K'\},\label{eq:Glashier_Graphical}
\end{equation}
see figure~\ref{fig:u-plane}, where we utilized the graphical interpretation
from \citep[(22.4)]{NIST:DLMF} and associate the denomination $\{\mathrm{s},\mathrm{c},\mathrm{d},\mathrm{n}\}$
with the four vertices of the quarter-periodicity rectangle. Eventually,
the quarter periods $K$ and $K'$ are the complete elliptic integrals
of the first kind, 
\begin{align}
K & \Def F\big(\tfrac{\pi}{2},k\big), & K' & \Def F\big(\tfrac{\pi}{2},k'\big),\label{eq:K-Kp-Def}
\end{align}
cf.~(\ref{eq:F(phi,k)}), with complementary modulus $k'$ fulfilling
$k^{2}+k'^{2}=1$. This elliptic parametrization will lead to substantial
simplifications of the results from \citep{Hucht16a}, as it firstly
eliminates the sign ambiguities of the square roots and secondly introduces
certain functions of the parameter $u$ that substantially simplify
the expressions.

\subsection{Coupling constants parametrization}

There are several possible ways to setup an elliptic parametrization
of the Onsager dispersion relation (\ref{eq:ODR}): one could set
$k=\hat{k}\Def z_{-}/t_{-}$ such that $0\leq\hat{k}<1$ holds in
the ordered phase. This choice is usually used in the literature \citep{IorgovLisovyy11,Baxter16}.
However, we will argue that many expressions become considerably simpler
if we instead use the modulus 
\begin{equation}
k\Def\frac{t_{-}}{z_{-}},\label{eq:k_Def}
\end{equation}
being the reciprocal of $\hat{k}$. It obeys $0\leq k<1$ in the disordered
phase $T>\Tc$ and $k>1$ in the ordered phase $T<\Tc$, similar to
the reduced inverse temperature $\beta/\beta_{\mathrm{c}}$. Note
that $K$ becomes complex for $k>1$, and we can replace it with its
real part, $K\mapsto\Re(K)=K+\ii K'$, in the complex analysis in
order to keep an un-tilted quarter-periodicity rectangle.

The anisotropy parameter $\eta$ can be introduced in several ways,
too. We choose the definition
\begin{equation}
\Jsn(2\eta)\Def\frac{1}{\ii t_{-}},\label{eq:eta_Def}
\end{equation}
such that $\eta$ is a purely imaginary point in the $u$-plane, with
$0\leq\Im(\eta)\leq K'/2$. This leads to the identities \bS\label{eqs:tz(eta)}
\begin{align}
t & =\frac{\Jsn\eta\Jdn\eta}{\ii\Jcn\eta}, & t_{+} & =\ii\Jcs(2\eta), & t_{-} & =\frac{1}{\ii}\Jns(2\eta),\label{eq:t(eta)}\\
z & =k\frac{\Jsn\eta\Jcn\eta}{\ii\Jdn\eta}, & z_{+} & =\frac{\ii}{k}\Jds(2\eta), & z_{-} & =\frac{1}{\ii k}\Jns(2\eta).\label{eq:z(eta)}
\end{align}
\eS{}For the dual couplings we find the corresponding expressions\bS
\begin{align}
{t^{*}}_{\!+} & =-\frac{t_{+}}{t_{-}}=\Jcn(2\eta), & {t^{*}}_{\!-} & =\frac{1}{t_{-}}=\ii\Jsn(2\eta),\label{eq:tdual_pm(eta)}\\
{z^{*}}_{\!+} & =-\frac{z_{+}}{z_{-}}=\Jdn(2\eta), & {z^{*}}_{\!-} & =\frac{1}{z_{-}}=\ii k\Jsn(2\eta),\label{eq:zdual_pm(eta)}
\end{align}
\eS{}which implies that we can express $t$, $z$, $t^{*}$ and $z^{*}$
through the Jacobi amplitude (\ref{eq:am_Def}), \bS\label{eqs:tz(am(eta))}
\begin{align}
t^{*} & =\ee^{\ii\Jam(2\eta)}, & t & =-\ii\tan\!\left[{\textstyle \frac{1}{2}}\Jam(2\eta)\right],\label{eq:t(am(eta))}\\
z & =\ee^{\ii\Jam(2\tilde{\eta})}, & z^{*} & =-\ii\tan\!\text{\ensuremath{\left[{\textstyle \frac{1}{2}}\Jam(2\tilde{\eta})\right]}},\label{eq:z(am(eta))}
\end{align}
\eS{}where $\tilde{u}$ denotes the swap transform (\ref{eq:flip_trans})
of a point $u$ in the complex $u$-plane, 
\begin{equation}
u\mapsto\tilde{u}\Def\fliptrans(u)={\textstyle \frac{1}{2}}\ii K'-u.\label{eq:flip_u}
\end{equation}
From (\ref{eq:F(phi,k)}) and the $t^{*}$ identity (\ref{eq:t(am(eta))})
we conclude that 
\begin{equation}
2\eta=F(-\ii\log t^{*},k)=\ii K'-F(-\ii\log z,k).\label{eq:eta_from_F}
\end{equation}
Note that in the isotropic case $t^{*}=z$ the point $\eta$ lies
symmetrically at 
\begin{equation}
\eta_{\mathrm{iso}}\Def{\textstyle \frac{1}{4}}\ii K'=\tilde{\eta}_{\mathrm{iso}},\label{eq:eta_iso}
\end{equation}
as $\fliptrans(\eta_{\mathrm{iso}})=\eta_{\mathrm{iso}}$. Therefore,
the transformation (\ref{eq:flip_u}) is a point reflection of the
complex $u$-plane at the point $\eta_{\mathrm{iso}}$.

\subsection{Eigenvalues parametrization}

The relation of the eigenvalues $\lambda$ and $\zeta$ (\ref{eqs:eigenvalues})
to $u$ is also ambiguous. We follow the literature \citep{Baxter82,IorgovLisovyy11,Baxter16}
and define
\begin{equation}
\frac{1}{\sqrt{\lambda\zeta^{\pm1}}}=\ee^{-\frac{\gamma\pm\ii\varphi}{2}}\Def\sqrt{k}\Jsn(u\mp\eta),\label{eq:sn(u_pm_eta)_Def}
\end{equation}
 such that 
\begin{align}
\lambda & =\ee^{\gamma}=\frac{1}{k\Jsn(u+\eta)\Jsn(u-\eta)}, & \zeta & =\ee^{\ii\varphi}=\frac{\Jsn(u+\eta)}{\Jsn(u-\eta)}.\label{eq:lambda-zeta-sn(u-pm-eta)}
\end{align}
The real and positive eigenvalues $\lambda_{\mu}$ correspond to values
$u_{\mu}$ on the real axis (for $\lambda_{\mu}>1$) as well as on
the line $\Im(u_{\mu})=\ii K'$ (for $\lambda_{\mu}<1$). With this
definition of $\lambda$, the constants $u_{\mathrm{p}}$ and $\lambda_{\mathrm{p}}$
from (\ref{eq:Glashier_Graphical}) and (\ref{eq:lambda_nscd_Def})
fulfill $\lambda_{\mathrm{p}}=\lambda(u_{\mathrm{p}})$ and represent
the upper and lower bound of the spectrum of $\mat T$ and $\mat T^{-1}$
both above and below $\Tc$. In appendix \ref{sec:Elliptic-Identities}
we give a subset of the large number of identities which can be derived
using elliptic functions identities \citep{Lawden89,NIST:DLMF}. Especially,
we see from (\ref{eqs:lambda(u)-zeta(u)}) that the eigenvalues $\lambda$
and $\zeta$ are exchanged under the transformation $\fliptrans$
according to $\fliptrans[(\lambda,\zeta)]=(\zeta,\lambda)$.

\begin{figure}
\begin{centering}
\includegraphics[width=0.95\textwidth]{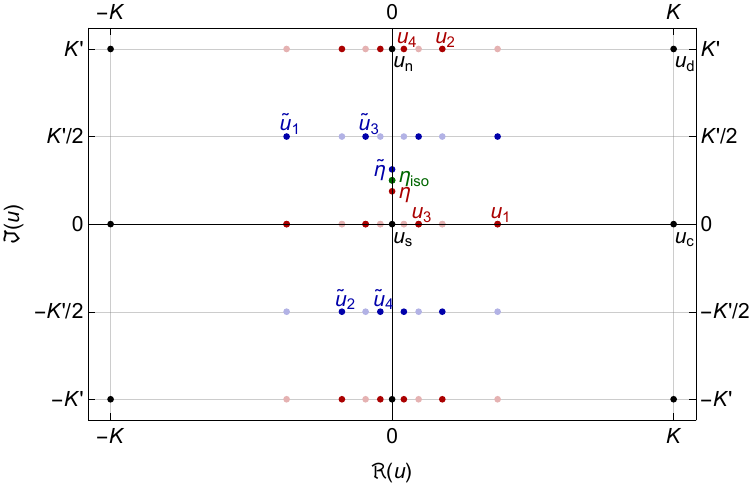}
\par\end{centering}
\caption{\linespread{1.2}\selectfont{}Structure of the complex elliptic $u$-plane
for $M=4$, paramagnetic temperature $k=0.95$, and anisotropy $\eta=\frac{3}{4}\eta_{\mathrm{iso}}$.
The eigenvalues $\lambda_{\mu}$ translate to the points $u_{\mu}$
(red), while the swap-transformed points $\tilde{u}_{\mu}$ (blue)
represent the eigenvalues $\zeta_{\mu}$ and lie point-symmetric w.r.t.~the
point $\eta_{\mathrm{iso}}$ (green) from (\ref{eq:eta_iso}). The
corners (\ref{eq:Glashier_Graphical}) of the quarter-periodicity
rectangle are shown in black. The additional eigenvalues $\lambda_{\mu}^{-1}$
at positions $\protect\ui_{\mu}\protect\Def u_{\mu}+\protect\ii K'$
(\ref{eq:inv_trans}) and the corresponding eigenvalues $\zeta_{\mu}^{-1}$
are shown in light colors. \label{fig:u-plane}}
\end{figure}
Besides the corners of the quarter-periodicity rectangle (\ref{eq:Glashier_Graphical}),
and the eigenvalues $u_{\mu}$, where the CP (\ref{eqs:CP+}) is zero,
there are four other important points in the complex $u$-plane, namely
the positions of the simple poles and simple zeroes of $\lambda(u)$
and $\zeta(u)$. Because the Jacobi $\Jsn u$ has a simple zero at
$u=0$ and a simple pole at $u=\ii K'$, we find \bS\label{eqs:counter_poles}
\begin{align}
u_{\lambda\to\infty,\zeta\to\infty} & =+\eta, & u_{\lambda\to0,\zeta\to\infty} & =+\ii K'-\eta,\label{eq:counter_poles_1}\\
u_{\lambda\to\infty,\zeta\to0} & =-\eta, & u_{\lambda\to0,\zeta\to0} & =-\ii K'+\eta.\label{eq:counter_poles_2}
\end{align}
\eS{}These four points will be used later in the analysis of the
complex structure of the relevant integrals. 

While $K$ and $K'$ are the quarter periods of the Jacobi elliptic
functions (\ref{eq:sncndn}), all relevant parameters and variables
of the considered system, such as $t,z,\lambda,\zeta$, can be written
as meromorphic functions of $2u$ and $2\eta$, see (\ref{eqs:tz(am(eta))}),
(\ref{eqs:lambda_with_2u}), and (\ref{eqs:zeta_with_2u}). They are
therefore double periodic functions with quarter periods $K/2$ and
$K'/2$ and \emph{half periods} $K$ and $K'$, and can be depicted
on the periodicity rectangle going from $-K-\ii K'$ to $K+\ii K'$,
as shown in figure~\ref{fig:u-plane}.

Finally, we remark that the CP zeroes show up in pairs $(u_{\mu},-\bar{u}_{\mu})$,
lying symmetrically with respect to the imaginary axis, see figure~\ref{fig:u-plane}.
The reason for this symmetry is the meromorphicity of the Jacobi elliptic
functions, and consequently of $\lambda(u)$ and $\zeta(u)$ from
(\ref{eq:lambda-zeta-sn(u-pm-eta)}), in combination with the fact
that both $\lambda$ and $\zeta$ are real for purely imaginary $u$.
Therefore, $\lambda$ and $\zeta$ transform according to $f(u)=\overline{f(-\bar{u})}$
under conjugation. This symmetry is sometimes called \textit{para-even}
in the literature, and $-\bar{u}$ is denoted the \textit{para-conjugate}
of $u$ \citep[6.29]{BultheelBarel}. We could restore the usual symmetry
along the real axis $f(u)=\overline{f(\bar{u})}$, valid for meromorphic
functions $f(u)$ that are real for real $u$, by rotating the complex
plane by $90^{\circ}$ using the Jacobi imaginary transform \citep[§22.6(iv)]{NIST:DLMF}.
This would however break the simple relation to the Jacobi amplitude
introduced in the next section. Alternatively, one could also replace
$u$ by $\ii u$ in all expressions as done in \citep{Baxter82},
but we will keep the present definition.

\subsection{The Jacobi amplitudes $\omega$ and $\theta$}

It turns out in the next chapter that the Jacobi amplitude $\omega$
of $2u$, as well as its imaginary swap  transform $\theta$, 
\begin{align}
\omega & \Def\Jam(2u), & \theta & \Def\ii\Jam(2\tilde{u}),\label{eq:omega_Def}
\end{align}
with $\tilde{u}$ from (\ref{eq:flip_u}), will play quite important
roles in the following\footnote{Note that \textit{Mathematica} \citep{MMA12} can correctly handle
the complex Jacobi amplitude only since version 12.1.}. They satisfy the identities\bS\label{eq:omega_identities}
\begin{align}
\Jsn(2u) & =\sin\omega, & \Jsn(2\tilde{u}) & =\frac{1}{\ii}\sinh\theta=-\frac{1}{k}\Jns(2u),\label{eq:omega_identities-sn}\\
\Jcn(2u) & =\cos\omega, & \Jcn(2\tilde{u}) & =\cosh\theta=\frac{\ii}{k}\Jds(2u),\label{eq:omega_identities-cn}\\
\Jdn(2u) & =-\coth\theta=\ii\Jcs(2\tilde{u}), & \Jdn(2\tilde{u}) & =\ii\cot\omega=\ii\Jcs(2u),\label{eq:omega_identities-dn}
\end{align}
\eS{}as well as 
\begin{align}
\tan({\textstyle \frac{1}{2}}\omega) & =\frac{\Jsn u\Jdn u}{\Jcn u}, & \ee^{\pm\theta} & =\left(k\frac{\Jsn u\Jcn u}{\ii\Jdn u}\right)^{\pm1},\label{eq:am_id_3_4}
\end{align}
and fulfill the symmetric relation
\begin{equation}
\ii k\sin\omega\sinh\theta=1.\label{eq:sin_w_sin_wt}
\end{equation}

In summary, we have defined a double periodic complex manifold with
the topology of a torus as originally proposed by Baxter \citep{Baxter82},
to describe the two-dimensional Ising model on the rectangle. This
complex manifold will be called the $u$-plane and is sketched in
figure~\ref{fig:u-plane}. The aspect ratio of the torus is temperature-dependent
and is encoded in the elliptic modulus $k$, while the coupling anisotropy
is described by a point $\eta$ on the torus. The $M$ eigenvalues
$\lambda_{\mu}$ and $\zeta_{\mu}$ correspond to points $u_{\mu}$
and $\tilde{u}_{\mu}$, respectively, on the torus. With this set
of definitions, we now can return to the characteristic polynomial
from section~\ref{subsec:Onsager-dispersion-relation}. 

\section{Results}

\subsection{Characteristic polynomials }

\label{sec:CharPoly}The characteristic polynomial $\P{\lambda_{+}}(\lambda_{+})$
of the matrix $\mat T_{+}$ was defined in chapter VI of \citep{Hucht16a}.
It was used to characterize the spectrum of $\mat T_{+}$ and to derive
the finite-size scaling limit in \citep{Hucht16b}. For the transfer
matrix $\mat T_{+}$ with eigenvalues $\lambda_{+,\mu}$, equation
(\ref{eq:CP+_2}) can be simplified to%
\begin{equation}
\P{\lambda_{+}}(\lambda_{+})=(1-{t^{*}}^{2})\left(\frac{t_{-}z_{-}}{2}\right)^{\negmedspace M}\left[\cos(M\varphi)+\frac{t_{+}z_{-}\cos\varphi-t_{-}z_{+}}{z_{-}\sin\varphi}\sin(M\varphi)\right],\label{eq:CP+-1}
\end{equation}
cf.~\cI{45}, where the angle\footnote{For the correct determination of the sign of $\varphi$ without elliptic
parametrization, see chapter VI in \citep{Hucht16a}.} $\varphi$ is given by the Onsager dispersion relation (\ref{eq:ODR}).
Using the elliptic parametrization from the last section, and especially
the Jacobi amplitude $\omega$ from (\ref{eq:omega_Def}), we find,
using (\ref{eqs:tz(eta)}) and (\ref{eqs:zeta_with_2u}), the surprisingly
simple expressions \bS\label{eq:CP+(u)} 
\begin{align}
\P{\lambda_{+}}(\lambda_{+}) & =(1-{t^{*}}^{2})\left(\frac{t_{-}z_{-}}{2}\right)^{\negmedspace M}\big[\cos(M\varphi)-\Jcs(2u)\sin(M\varphi)\big]\label{eq:CP+(u)_1}\\
 & =(1-{t^{*}}^{2})\left(\frac{t_{-}z_{-}}{2}\right)^{\negmedspace M}\frac{\sin(M\varphi-\omega)}{\sin(-\omega)},\label{eq:CP+(u)_2}
\end{align}
\eS{}as $\Jcs(2u)=\cot\omega$ by (\ref{eq:omega_identities-dn}).
Here, $\lambda_{+}$ and $\varphi$ depend on $u$ according to (\ref{eq:lambda-zeta-sn(u-pm-eta)}),
(\ref{eq:lambda(u)-zeta(u)}), or (\ref{eqs:lambda_with_2u}) and
(\ref{eqs:zeta_with_2u}). As a consequence, the eigenvalues $\lambda_{+,\mu}$
of $\mat T_{+}$ fulfill the simple condition $M\varphi_{\mu}=\omega_{\mu}$.
These simplifications demonstrate the power of the elliptic parametrization
in the chosen form, and the introduced Jacobi amplitude $\omega$
(\ref{eq:omega_Def}) turns out to be a phase shift in $\varphi$-space,
describing the open BCs in $\M$ direction. 

Due to the linear CP identity 
\begin{equation}
P_{c\boldsymbol{a}+b}(cx+b)=\prod_{\mu=1}^{M}(cx+b-ca_{\mu}-b)=c^{M}\P a(x),\label{eq:CP_Identity}
\end{equation}
we can rewrite (\ref{eq:CP+(u)_2}) in terms of the new variable 
\begin{equation}
\chi\Def2(\zeta_{+}+1)=4\cos^{2}({\textstyle \frac{1}{2}}\varphi)=\frac{2}{t_{-}z_{-}}(t_{+}z_{+}+t_{-}z_{-}-\lambda_{+}),\label{eq:chi_Def}
\end{equation}
eliminating the factor $(t_{-}z_{-}/2)^{M}$, and find the corresponding
CP
\begin{equation}
\P{\chi}(\chi)=(1-{t^{*}}^{2})\,\frac{\sin(M\varphi-\omega)}{\sin(-\omega)}.\label{eq:CP_chi}
\end{equation}
Comparing (\ref{eq:chi_Def}) and (\ref{eq:Tp}), we see that $\P{\chi}(\chi)$
is the characteristic polynomial of the matrix $\mat C$, which therefore
has the eigenvalues $\chi_{\mu}$.

\subsection{Common eigenvectors}

With the help of the introduced elliptic parametrization, the matrix
$\mat x$ of orthonormal common eigenvectors\footnote{The eigenvectors $\vec{x}_{\mu}$ are row vectors in $\mat x$, i.\,e.,
$\vec{x}_{\mu}=(\mat x)_{\mu}$.} of $\mat T_{\pm}$, $\mat T$ and $\mat C$, defined in (\ref{eq:eigensystem})
and originally given in \cI{50}, can be considerably simplified,
too. Using the projectors $\mat S^{\pm}$ from (\ref{eq:Spm}), we
can split $\mat x$ into an even part $\mat x^{+}$ and an odd part
$\mat x^{-}$ according to $\mat x^{\pm}\Def\mat x\mat S^{\pm}$,
such that $\mat x=\mat x^{+}+\mat x^{-}$ and $\mat x^{\pm}=\pm\mat x^{\pm}\mat S$,
i.\,e., $\mat x^{+}$ ($\mat x^{-}$) contains the symmetric (skew-symmetric)
parts of the eigenvectors. We get \bS
\begin{align}
\mat x^{+} & =\tfrac{1}{\sqrt{2}}\mat D^{\frac{1}{2}}\left[\ee^{-\frac{1}{2}(\theta_{\mu}-\psi_{\mu}+\frac{\ii\pi}{2})}\,\frac{\cos({\textstyle \frac{m}{2}}\varphi_{\mu})}{\cos({\textstyle \frac{1}{2}}\varphi_{\mu})}\right]_{\mu=1,\,m\,\text{odd}}^{M},\label{eq:x^+}\\
\mat x^{-} & =\tfrac{1}{\sqrt{2}}\mat D^{\frac{1}{2}}\left[\ee^{+\frac{1}{2}(\theta_{\mu}-\psi_{\mu}+\frac{\ii\pi}{2})}\,\frac{\sin({\textstyle \frac{m}{2}}\varphi_{\mu})}{\sin({\textstyle \frac{1}{2}}\varphi_{\mu})}\right]_{\mu=1,\,m\,\text{odd}}^{M},\label{eq:x^-}
\end{align}
\eS{}where $m$ runs over the odd integers between $-M$ and $M$.
The angle $\psi$ is defined through 
\begin{equation}
\ee^{\frac{1}{2}(\theta-\psi)}\Def\sqrt{\ii z}\,\frac{\Jcn u}{\Jcn\eta},\label{eq:psi_Def}
\end{equation}
with $\theta$ from (\ref{eq:omega_Def}), cf.~(\ref{eq:am_id_3_4})
and (\ref{eq:id1-tan}), and fulfills 
\begin{equation}
\ee^{-\psi}=-\ii\zeta^{*}=-\tan({\textstyle \frac{1}{2}}\varphi),\label{eq:psi_1}
\end{equation}
see also (\ref{eq:prod_cn}). Using the CP $\P{\lambda_{+}}(\lambda_{+})$
from (\ref{eq:CP+(u)}), or alternatively $\P{\chi}(\chi)$ from (\ref{eq:CP_chi}),
the diagonal normalization matrix $\mat D$ can be substantially simplified
from the expression given in \cI{50}, with the result\bS\label{eq:D}
\begin{align}
(\mat D)_{\mu\mu} & =-\frac{t^{*}}{z_{-}}\left(\frac{t_{-}z_{-}}{2}\right)^{\negmedspace M-1}\frac{1}{\P{\lambda_{+}}'(\lambda_{+,\mu})}\label{eq:D_lambda_+}\\
 & =\frac{t^{*}}{z_{-}}\frac{1}{\P{\chi}'(\chi_{\mu})}.\label{eq:D_xi_+}
\end{align}
\eS{}Here we used a relation between the normalization of the eigenvectors
of a tridiagonal matrix and the derivative $P'$ of its CP \citep[chapter 7.9]{Parlett80}.
With these simplifications and the diagonal eigenvalue matrix $(\mat{\Lambda})_{\mu\mu}=\lambda_{\mu}$,
we can rewrite (\ref{eq:M_def}) as 
\begin{equation}
\mat M=\mat{\Lambda}^{\negmedspace L/2}\,\mat x^{+}+\mat{\Lambda}^{\negmedspace-L/2}\,\mat x^{-},\label{eq:M_x^+_x^-}
\end{equation}
and the partition function (\ref{eq:det(T2)}) becomes $Z=Z_{0}\det\mat M$.
Inserting the definition of $\mat x^{\pm}$ into (\ref{eq:M_x^+_x^-}),
we get the result
\begin{equation}
\hat{\mat W}\Def\frac{1}{\sqrt{2}}\left[\ee^{\frac{1}{2}(L\gamma_{\mu}-\theta_{\mu}+\psi_{\mu}-\frac{\ii\pi}{2})}\,\frac{\cos({\textstyle \frac{m}{2}}\varphi_{\mu})}{\cos({\textstyle \frac{1}{2}}\varphi_{\mu})}+\ee^{-\frac{1}{2}(L\gamma_{\mu}-\theta_{\mu}+\psi_{\mu}-\frac{\ii\pi}{2})}\,\frac{\sin(\frac{m}{2}\varphi_{\mu})}{\sin({\textstyle \frac{1}{2}}\varphi_{\mu})}\right]_{\mu=1,\,m\,\text{odd}}^{M},\label{eq:W_hat}
\end{equation}
such that (\ref{eq:MM}) becomes $\T{\mat M}\mat M=\hat{\mat W}\T{\vphantom{\mat W}}\mat D\hat{\mat W}.$ 

As shown in \citep{Hucht16a}, the matrix $\hat{\mat W}$ is a generalized
Vandermonde matrix, such that a change of the base leaves the value
of the determinant invariant. Hence we can transform from the trigonometric
base to the simpler power base and get the corresponding matrix
\begin{equation}
\mat W\Def\left[\ee^{H(m)(L\gamma_{\mu}-\theta_{\mu}+\psi_{\mu})}\,\chi_{\mu}^{\frac{1}{2}(|m|-1)}\right]_{\mu=1,\,m\,\text{odd}}^{M},\label{eq:W}
\end{equation}
with the Heaviside step function $H$, where we have used the variable
$\chi$ introduced in (\ref{eq:chi_Def}), and factored out the exponential
for $m<0$ and moved it to the diagonal matrix 
\begin{equation}
\mat F\Def\left[\delta_{\mu\nu}\,\ee^{-\frac{1}{2}(L\gamma_{\mu}-\theta_{\mu}+\psi_{\mu})}\right]_{\mu,\nu=1}^{M},\label{eq:F_mumu}
\end{equation}
such that $\det\hat{\mat W}=\det\mat F\det\mat W$. As $\det\mat F=t^{-\frac{L}{2}}$
by (\ref{eq:psi_Def}), (\ref{eq:det_T}) and (\ref{eq:prod_cn}),
the resulting squared partition function now reads
\begin{equation}
Z^{2}=Z_{0}^{2}\,t^{-L}\det(\T{\mat W}\mat D\mat W).\label{eq:Zq}
\end{equation}
Up to here, the calculation was similar to \citep{Hucht16a}, with
one major difference: the occurrence of the terms $1/\P{\chi}'(\chi_{\mu})$
in the diagonal matrix $\mat D$ (\ref{eq:D_xi_+}) will enable us
to use a relation between characteristic polynomials, their associated
Vandermonde matrices and certain Hankel matrices.

\subsection{Hankelization}

The next significant simplification is obtained by utilizing a generalization
of the well known relation of the characteristic polynomial $\P x(x)$
and the related Vandermonde matrix%
\begin{equation}
\mat V_{\!\boldsymbol{x}}\Def\left[\vphantom{A_{2}^{2}}x_{\mu}^{j}\right]_{\mu=1,\,j=0}^{M\hspace{2ex}M-1}\label{eq:V_x}
\end{equation}
to a certain Hankel matrix $\mat H_{\boldsymbol{x}}$ \citep{HeinigRost,LutherRost04},
also known as 'Vandermonde factorization of a Hankel matrix'. Let
\begin{equation}
\P x(x)=\prod_{\mu=1}^{M}(x-x_{\mu})=\sum_{n=0}^{M}b_{n}x^{n},\label{eq:P_x(x)}
\end{equation}
where $b_{M}=1$ by construction, as well as 
\begin{align}
\mat D_{\boldsymbol{x}} & \Def\left[\frac{\delta_{\mu\nu}}{\P x'(x_{\mu})}\right]_{\mu,\nu=1}^{M}, & \mat H_{\boldsymbol{x}}^{-1} & \Def\left[\vphantom{A_{2}^{2}}b_{i+j+1}\right]_{i,j=0}^{M-1},\label{eq:D_x,H_x}
\end{align}
then 
\begin{align}
\T{\mat V}_{\!\boldsymbol{x}}\,\mat D_{\boldsymbol{x}}\mat V_{\!\boldsymbol{x}} & =\mat H_{\boldsymbol{x}}, & {\det}^{2}\,\mat V_{\!\boldsymbol{x}}\det\mat D_{\boldsymbol{x}} & =1.\label{eq:Vandermonde_Hankel_Identity}
\end{align}
While $\mat H_{\boldsymbol{x}}^{-1}$ is upper anti-triangular, $\mat H_{\boldsymbol{x}}$
itself is lower anti-triangular, e.\,g. for $M=4$,
\begin{equation}
\setlength{\arraycolsep}{5pt}\mat H_{\boldsymbol{x}}^{-1}=\begin{pmatrix}b_{1} & b_{2} & b_{3} & 1\\
b_{2} & b_{3} & 1 & 0\\
b_{3} & 1 & 0 & 0\\
1 & 0 & 0 & 0
\end{pmatrix}\quad\Rightarrow\quad\mat H_{\boldsymbol{x}}=\begin{pmatrix}0 & 0 & 0 & 1\\
0 & 0 & 1 & \tilde{b}_{5}\\
0 & 1 & \tilde{b}_{5} & \tilde{b}_{6}\\
1 & \tilde{b}_{5} & \tilde{b}_{6} & \tilde{b}_{7}
\end{pmatrix},\label{eq:H_x_example}
\end{equation}
with\footnote{As a side note, both $\tilde{b}_{n}=s_{(n-M)}(\boldsymbol{x})$ and
$b_{n}=(-1)^{n}s_{\boldsymbol{1}_{M-n}}(\boldsymbol{x})$ are Schur
polynomials.} $\tilde{b}_{n}=\sum_{\mu}x_{\mu}^{n-1}/\P x'(x_{\mu})$. 

A direct computation (see appendix \ref{sec:Block-Hankel-matrix})
shows that a similar identity holds for the generalized Vandermonde
matrix $\mat W$ (\ref{eq:W}) in conjunction with $\mat D$ from
(\ref{eq:D_xi_+}), namely
\begin{equation}
\T{\mat W}\mat D\mat W=\begin{bmatrix}\mat 0 & \mat S\hat{\mat H}_{1}\\
\hat{\mat H}_{1}\mat S & \hat{\mat H}_{2}
\end{bmatrix}\label{eq:generalized_V_H_Identity}
\end{equation}
with the $\nicefrac{M}{2}\times\nicefrac{M}{2}$ Hankel matrices
\begin{equation}
\hat{\mat H}_{\Delta}\Def\left[\sum_{\mu=1}^{M}\frac{t^{*}}{z_{-}}\frac{\ee^{\Delta(L\gamma_{\mu}-\theta_{\mu}+\psi_{\mu})}}{\P{\chi}'(\chi_{\mu})}\,\chi_{\mu}^{i+j}\right]_{i,j=0}^{\frac{M}{2}-1}.\label{eq:H_Delta}
\end{equation}
Note that (\ref{eq:generalized_V_H_Identity}) is block lower anti-triangular
similar to $\mat H_{\boldsymbol{x}}$ in (\ref{eq:H_x_example}),
as $\hat{\mat H}_{0}=\mat 0$. The important consequence of this result
is the generalized determinant identity
\begin{equation}
\det(\T{\mat W}\mat D\mat W)={\det}^{2}\,\mat W\det\mat D={\det}^{2}\,\hat{\mat H}_{1},\label{eq:det_id_H1}
\end{equation}
which leads to the surprising result that the Ising partition function
can be mapped to a Hankel determinant. Inserting the simplifications
from above, and defining $\mat H\Def2\ii z_{-}\hat{\mat H}_{1},$
we can draw the square root in (\ref{eq:Zq}) and get the compact
exact expression for the partition function of the square lattice
Ising model on the rectangle,\bS
\begin{align}
Z & =Z_{1}\det\mat H, & Z_{1} & \Def t^{-\frac{L}{2}}z^{\frac{M}{2}}\left(-\frac{2}{z_{-}}\right)^{\negmedspace\frac{LM}{2}},\label{eq:Z-Z1}
\end{align}
where the $\nicefrac{M}{2}\times\nicefrac{M}{2}$ Hankel matrix $\mat H=\left[\vphantom{A_{2}^{2}}h_{i+j+1}\right]_{i,j=0}^{M/2-1}$
has the matrix elements 
\begin{equation}
h_{n}\Def\sum_{\mu=1}^{M}\frac{2\ii t^{*}\,\ee^{L\gamma_{\mu}-\theta_{\mu}+\psi_{\mu}}}{\P{\chi}'(\chi_{\mu})}\,\chi_{\mu}^{n-1},\label{eq:h_n_1}
\end{equation}
\eS{}with $n=1,\ldots,M-1$. This expression represents a significant
simplification with respect to the result from \citep{Hucht16a}.
However, in the next section we will proceed further by rewriting
the sum over $\mu$ as a complex contour integral, inserting the known
formula for $\P{\chi}(\chi)$ from (\ref{eq:CP_chi}).

\subsection{Contour integral representation }

The matrix elements $h_{n}$ of the Hankel matrix $\mat H$ (\ref{eq:h_n_1})
can be evaluated using complex contour integration, and the characteristic
polynomial $\P{\chi}(\chi)$ plays a crucial role in this calculation.
In principle we use Cauchy's residual theorem and calculate the sum
over $\mu$ in $h_{n}$ as a contour integral over a suitable contour
$C$ around the points $u_{\mu}$ in the complex $u$-plane,\bS%
\begin{align}
h_{n} & =\frac{1}{2\pi\ii}\oint_{C}\frac{2\ii t^{*}\,\ee^{L\gamma-\theta+\psi}}{\P{\chi}'(\chi)}\,\chi^{n-1}\,\frac{\partial\log\P{\chi}(\chi)}{\partial\chi}\frac{\partial\chi}{\partial u}\,\dd u\label{eq:H_n_2}\\
 & =\frac{1}{2\pi\ii}\oint_{C}\frac{2\ii t^{*}\,\ee^{L\gamma-\theta+\psi}}{\P{\chi}(\chi)}\,\chi^{n-1}\,\frac{\partial\chi}{\partial u}\,\dd u,\label{eq:H_n_3}
\end{align}
\eS{}where, most importantly, the derivative $\P{\chi}'(\chi)$ cancels
out. 

While the CP $\P{\chi}(\chi)$ has two sets of zeroes $u_{\mu}$ and
$\ui_{\mu}$ corresponding to the eigenvalues $\lambda_{\mu}$ and
$\lambda_{\mu}^{-1}$, cf.~(\ref{eq:inv_trans}), the contour $C$
only encloses the zeroes $u_{\mu}$, see Figure \ref{fig:u-plane}.
We can easily remove the additional zeroes $\ui_{\mu}$ by employing
a factorization analog to (\ref{eq:CP+_Factor}), 
\begin{equation}
\P{\chi}(\chi)=\frac{1-{t^{*}}^{2}}{2\ii\sin\omega}\big[1-\ee^{\ii(M\varphi-\omega)}\big]\big[1+\ee^{-\ii(M\varphi-\omega)}\big].\label{eq:CP_chi_factor}
\end{equation}
As the first (second) bracket vanishes at $u_{\mu}$ ($\ui_{\mu}$),
we can drop the additional zeroes $\ui_{\mu}$ by replacing the last
term with its value at the zeroes $u_{\mu}$, where $M\varphi-\omega=0$,
to get
\begin{equation}
h_{n}=\frac{1}{2\pi\ii}\oint_{C}\frac{\ee^{L\gamma-\theta+\psi}}{1-\ee^{\ii(M\varphi-\omega)}}\,t_{-}\sin\omega\,\chi^{n-1}\,\frac{\partial\chi}{\partial u}\,\dd u.\label{eq:H_n_4}
\end{equation}
In the next step we use (\ref{eq:psi_1}), (\ref{eq:dgamma_du}) and
(\ref{eq:dgamma_dphi-dchi_du}) to eliminate $\psi$ and $\sin\omega$.
Furthermore, we can move the zeroes of the numerator to the line $\Im(u)=\frac{1}{2}K'$
without changing the integral by adding one to the numerator, such
that the numerator $1-\ee^{L\gamma-\theta}=\fliptrans[1-\ee^{\ii(M\varphi-\omega)}]$
is precisely the swap transform (\ref{eq:flip_trans}) of the denominator,
with the result
\begin{equation}
h_{n}=\frac{1}{2\pi\ii}\oint_{C}\frac{1-\ee^{L\gamma-\theta}}{1-\ee^{\ii(M\varphi-\omega)}}\,\chi^{n}\,\frac{\partial\gamma}{\partial u}\,\dd u.\label{eq:H_n_5}
\end{equation}
Due to the CP property (\ref{eq:CP_Identity}) and the Vandermonde
property of (\ref{eq:W_hat}), the determinant of (\ref{eq:H_n_4})
is invariant under a translation $\chi\mapsto\chi+c$. This freedom
is used in (\ref{eq:H_n_5}), as $\chi$ was defined in (\ref{eq:chi_Def})
in order to obey $\chi=2\cot(\frac{1}{2}\varphi)\sin\varphi$. 

\begin{figure}
\begin{centering}
\includegraphics[width=0.49\columnwidth]{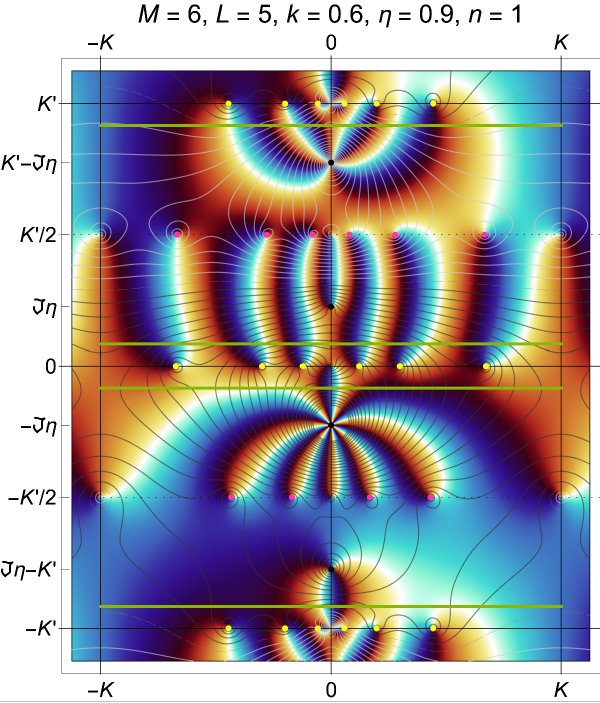}~\includegraphics[width=0.49\columnwidth]{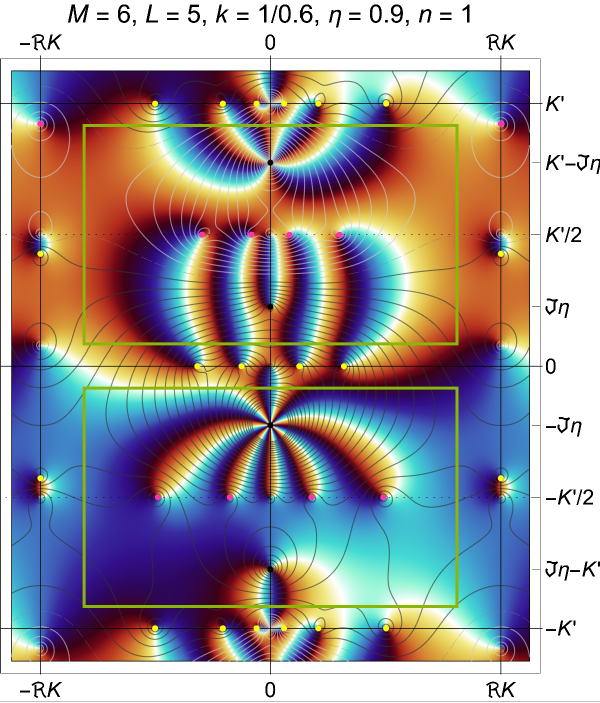}
\par\end{centering}
\caption{\linespread{1.2}\selectfont{}Complex structure of the integrand (\ref{eq:H_n_5})
for $M=6$, $L=5$, $n=1$ and anisotropy $\eta=0.9\,\eta_{\mathrm{iso}}$,
above ($k=0.6$, left) and below ($k^{-1}=0.6$, right) the critical
point. Yellow: $M$ zeroes of the denominator, i.\,e., eigenvalues
$\lambda_{\mu}$ of $\protect\mat T$, and their para-conjugate. Pink:
$L$ zeroes of the numerator and their para-conjugate. Black: multiple
zeroes/poles from (\ref{eq:counter_poles_1}), with pole order $\{-4,1,7,2\}$
from top to bottom in this case. Green: possible integration paths.
The contour lines are at constant modulus, with a dashed gray line
at 1, and light (dark) gray lines at powers of 2 below (above) 1.
The complex phase is color coded, being \{white, red, black, blue\}
at $\{1,\protect\ii,-1,-\protect\ii\}$, such that white turns to
red (blue) at zeroes (poles) under ccw.~rotation. Note that $K$
becomes complex for $k>1$. \label{fig:Integrand}}
\end{figure}
\begin{figure}
\begin{centering}
\includegraphics[width=0.49\columnwidth]{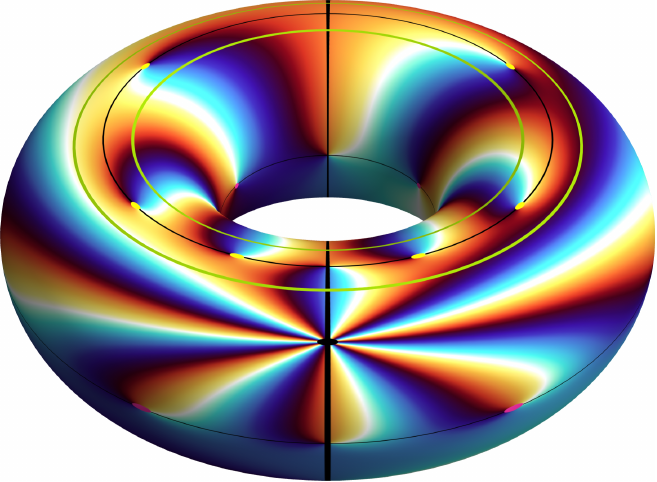}\hfill{}\includegraphics[width=0.49\columnwidth]{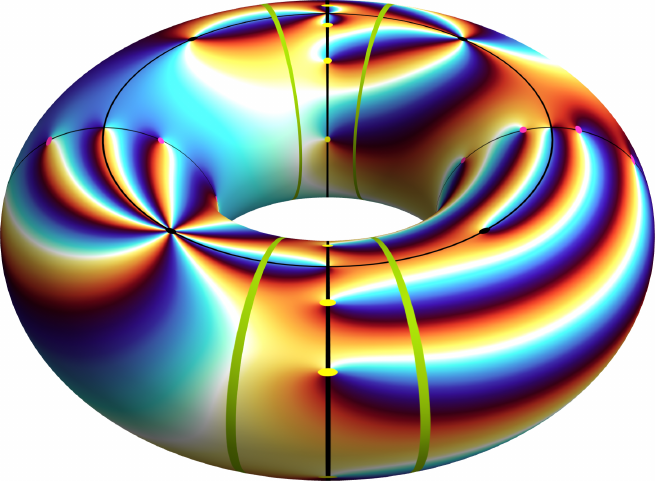}
\par\end{centering}
\caption{\linespread{1.2}\selectfont{}Complex structure of the integrand (\ref{eq:H_n_5})
from figure \ref{fig:Integrand} (left), mapped onto a torus. These
pictures might (or might not) be helpful to better understand the
setup. Left: the upper circle is the real axis, with the odd zeroes
of the denominator (yellow), the outer circle has $\Im(u)=-K'/2$,
with the even zeroes of the numerator (pink), and the black point
in front is at $-\eta$. The remaining zeroes are inside and at the
bottom. Right: The top circle is the imaginary axis $\Re(u)=0$, the
front ring is the real axis. This representation corresponds to the
original directions of the Ising model: the $\protect\M$ direction
goes along the $2M$ denominator zeroes (yellow) from front to back,
the $\protect\L$ direction goes along the $2L$ numerator zeroes
(pink) from left to right, see also table~\ref{tab:Quantities}.\label{fig:torus}}
\end{figure}
The resulting integrand is shown in figures \ref{fig:Integrand}.
At the four points $\{u_{0,\infty},u_{\infty,\infty},u_{\infty,0},u_{0,0}\}$
(\ref{eqs:counter_poles}) (black dots), the pole orders\footnote{$n$-fold poles (zeroes) have pole order $n$ ($-n$).}
$\{n+1-M,n+1+L-M,n+1+L,n+1\}$ are positive for certain $n\in\{1,\ldots,M-1\}$.
As the additional zeroes at $\ui_{\mu}$ are removed, we can deform
and simplify the integration contour $C$ to the four straight lines
(green), which pairwise enclose the CP zeroes (yellow), or equivalently,
the four points (\ref{eqs:counter_poles}) (black dots). Due to the
double periodic complex plane, the integration paths at $\Re(u)=\pm K$
add up to zero. Note that in the ordered phase below $\Tc$ the smallest
real zero $u_{1}$ reaches $\pm K$ and becomes complex, such that
the integration contour has to be modified as indicated. In figure~\ref{fig:torus}
the complex torus is depicted.

Summarizing the last steps, we have found an extremely compact representation
for the partition function of the anisotropic square lattice Ising
model on the rectangle as the determinant of a Hankel matrix $\mat H=\left[\vphantom{A_{2}^{2}}h_{i+j+1}\right]_{i,j=0}^{M/2-1}$,
\begin{align}
Z & =Z_{1}\det\mat H, & h_{n} & =\frac{1}{2\pi\ii}\oint_{C}\frac{1-\ee^{L\gamma-\theta}}{1-\ee^{\ii(M\varphi-\omega)}}\,\chi^{n}\,\frac{\partial\gamma}{\partial u}\,\dd u,\label{eq:Z-h_n_2}
\end{align}
with constant $Z_{1}$ from (\ref{eq:Z-Z1}). The $h_{n}$ are Fourier
coefficients (in the $\chi$ base) of a so-called symbol function
with Fisher-Hartwig type singularities \citep{FisherHartwig68,HartwigFisher69}
(see below), which is given by the ratio of two characteristic polynomials.
These polynomials are associated with the two directions as well as
the corresponding boundary conditions. As the considered system is
invariant under an exchange of the two directions, the two CPs are
directly related by the symmetry (\ref{eq:flip_trans}) of the underlying
square lattice Ising model.

Note that the $L$ zeroes $v_{\ell}$ of the numerator $1-\ee^{L\gamma-\theta}$
correspond to the $L$ eigenvalues of a hypothetical $L\times L$
transfer matrix $\tilde{\mat T}$ propagating in vertical direction,
which would have been used in an alternative rotated setup. Only in
the isotropic square system, where $L=M$ and $\K^{\L}=\K^{\M}$,
these zeroes coincide with the swap-transformed zeroes, $v_{\ell}=\fliptrans(u_{\mu})=\tilde{u}_{\mu}$.
Unfortunately, we were not able to simplify the integral (\ref{eq:Z-h_n_2})
in this symmetric case.

\subsection{From Hankel to Toeplitz}

The Hankel matrix (\ref{eq:H_n_5}) can be further transformed using
the identities derived by Basor \& Ehrhardt \citep{BasorEhrhardt02}.
Using theorem 2.3, the symbols $a$ and $b$, as well as the notation
$T_{M}(a)$, $H_{M}(a)$ and $H_{M}[b]$ from \citep{BasorEhrhardt02},
the determinant of $\mat H$, with elements (\ref{eq:H_n_5}), corresponds
to the Hankel moment determinant $\det H_{M}[b]$ and can be transformed
to the symmetric Toeplitz plus Hankel determinant $\det(T_{M}(a)+H_{M}(a))$,
with the symbol 
\begin{equation}
a(\zeta)=\ee^{-\psi}b(\zeta_{+})=\frac{1-\ee^{L\gamma-\theta}}{1-\ee^{\ii(M\varphi-\omega)}}\,\frac{\partial\gamma}{\ii\partial\varphi},\label{eq:symbols_a_b}
\end{equation}
and with the Fourier components \citep[eq.~(3)]{BasorEhrhardt02}\bS
\begin{align}
a_{n}=a_{-n} & =\frac{1}{2\pi}\oint_{C}a(\ee^{\ii\varphi})\,\ee^{-\ii n\varphi}\,\frac{\partial\varphi}{\partial u}\,\dd u\label{eq:symbol_a_1}\\
 & =\frac{1}{2\pi\ii}\oint_{C}\frac{1-\ee^{L\gamma-\theta}}{1-\ee^{\ii(M\varphi-\omega)}}\zeta^{-n}\,\frac{\partial\gamma}{\partial u}\,\dd u.\label{eq:symbol_a_2}
\end{align}
\eS{}Note that (\ref{eq:symbols_a_b}) transforms to its reciprocal,
$\fliptrans[a(\zeta)]=1/a(\zeta)$, under exchange of the two directions.
The advantage of this representation is the occurrence of the simpler
Fourier base $\zeta^{-n}$, which eliminates the poles in the integrand
stemming from $\zeta\to\infty$ at the points $u_{\infty,\infty}=\eta$
and $u_{0,\infty}=\ii K'-\eta$, provided $L<M$ (as $n\geq0$). We
can therefore simplify the integration contour to a curve around the
two remaining poles at $u_{\infty,0}=-\eta$ and $u_{0,0}=-\ii K'+\eta$.
However, for arbitrary $L$ and $M$ the pole order of the integrand
(\ref{eq:symbol_a_2}) at the four points (\ref{eqs:counter_poles})
is $\{-n+2-M,-n+L-M,n+L,n\}$ and can be positive in all four cases,
such that all four points must be enclosed by the contour.

Finally, from chapter 3 of \citep{BasorEhrhardt02} we borrow a clever
transformation from the symmetric Toeplitz plus Hankel representation
to a skew-symmetric $M\times M$ Toeplitz matrix $\hat{\mat T}$,
with the result \bS
\begin{align}
\hat{\mat T} & =\left[\sum_{\mu=1}^{M}\frac{2\ii t^{*}\,\ee^{L\gamma_{\mu}-\theta_{\mu}}}{\P{\chi}'(\chi_{\mu})\sin\varphi_{\mu}}\cot(\tfrac{1}{2}\varphi_{\mu})\sin\!\big[(i-j)\varphi_{\mu}\big]\right]_{i,j=0}^{M-1}\label{eq:T_skew_sum}\\
 & =\left[\frac{1}{\pi\ii}\oint_{C}\frac{1-\ee^{L\gamma-\theta}}{1-\ee^{\ii(M\varphi-\omega)}}\cot(\tfrac{1}{2}\varphi)\sin\!\big[(i-j)\varphi\big]\,\frac{\partial\gamma}{\partial u}\,\dd u\right]_{i,j=0}^{M-1},\label{eq:T_skew_int}
\end{align}
such that the Pfaffian of $\hat{\mat T}$ fulfills $\Pf\hat{\mat T}=\det\mat H$,
and therefore 
\begin{equation}
Z=Z_{1}\Pf\hat{\mat T}.\label{eq:Z_Pf_T}
\end{equation}
\eS{}As before, the integration contour $C$ can be freely deformed
as long as it enclosed all zeroes $u_{\mu}$ or, alternatively, the
four points (\ref{eqs:counter_poles}). 

\section{Discussion}

In this work, we showed that the partition function of the anisotropic
square lattice Ising model on the $L\times M$ rectangle with open
boundary conditions in both directions is given by the determinant
of a $\nicefrac{M}{2}\times\nicefrac{M}{2}$ Hankel matrix $\mat H$,
which equivalently can be written as the Pfaffian of a skew-symmetric
$M\times M$ Toeplitz matrix $\hat{\mat T}$. The $M-1$ independent
matrix elements of $\mat H$ or $\hat{\mat T}$ are Fourier coefficients
of a symbol function (\ref{eq:symbols_a_b}), which is given by the
ratio of two characteristic polynomials. These polynomials are associated
to the two directions $(\L,\M)$ of the system, encode the respective
boundary conditions, and are directly related through the symmetry
of the square lattice Ising model under exchange of the two directions. 

In the framework of the square lattice Ising model, Toeplitz matrices
and determinants are well known in the context of bulk spatial correlation
functions $\langle\sigma_{0,0}\sigma_{\ell,m}\rangle$ \citep{FisherHartwig68,McCoyWu73,DeiftItsKrasovsky13},
eventually leading to the spontaneous magnetization for $\ell^{2}+m^{2}\text{\ensuremath{\to\infty}}$.
Surprisingly, they now also appear in the exact expressions for the
partition function $Z$ of finite systems.

\CC{We explicitly point out that the simplifications with respect
to \citep{Hucht16a} derived in this work go far beyond ``rewriting
a determinant as another determinant'', mainly because (i) Hankel
and Toeplitz determinants are much simpler than regular determinants
and (ii) the associated symbol function (\ref{eq:symbols_a_b}) is
substantially simpler that the corresponding expressions from \citep{Hucht16a}.
}

The considered anisotropic Ising model with open BCs in both directions
is invariant under exchange of the two directions $\L$ and $\M$,
such that both the system dimensions and the coupling constants are
exchanged according to $L\stackrel{\fliptrans}{\leftrightarrow}M$
and $\K^{\L}\stackrel{\fliptrans}{\leftrightarrow}\K^{\M}$, see table~\ref{tab:Quantities}.
This  \emph{swap transformation} $\fliptrans$ (\ref{eq:flip_trans})
leads to the mapping of $z$ and $t$ to their respective duals, $(z,t)\stackrel{\fliptrans}{\mapsto}(t^{*},z^{*})$,
as well as to the exchange of the eigenvalues and angles according
to $\lambda\stackrel{\fliptrans}{\leftrightarrow}\zeta$, $\gamma\stackrel{\fliptrans}{\leftrightarrow}\ii\varphi$
and $\theta\stackrel{\fliptrans}{\leftrightarrow}\ii\omega$. In the
complex $u$-plane, it corresponds to a point reflection $(u,\eta)\stackrel{\fliptrans}{\mapsto}(\tilde{u},\tilde{\eta})$
of the complex torus at the point $\eta_{\mathrm{iso}}={\textstyle \frac{1}{4}}\ii K'$
from (\ref{eq:eta_iso}). The elliptic modulus $k$, however, is invariant
under this transformation. This symmetry must not be confused with
the related duality transformation $\mathcal{D}:(z,t^{*})\mapsto\mathcal{D}[(z,t^{*})]=(t,z^{*})$
of the bulk system, which maps a low-temperature system with spins
$\sigma_{\ell,m}$ at $k>1$ to a high-temperature system with the
plaquette spins $\tilde{\sigma}_{\ell,m}$ at $k\mapsto\tilde{k}=k^{-1}<1$,
see figure~\ref{fig:lattice}. Note that in systems with boundaries,
this transform is known to change the boundary conditions, e.\,g.,
from open to fixed.

\begin{table}[b]
\caption{Overview over the quantities used in this work. The transformation
$\protect\fliptrans$ (\ref{eq:flip_trans}) maps between the two
directions of the model. The right columns show the relation of our
notation to the work of Baxter \citep{Baxter16}, where the transfer
matrix propagated vertically. \label{tab:Quantities}}

\centering{}%
\begin{tabular}{l|>{\centering}p{0.2\columnwidth}>{\centering}p{0.2\columnwidth}|>{\centering}p{0.1\columnwidth}>{\centering}p{0.1\columnwidth}}
direction & $\L$ & $\M$ & $\M$ \citep{Baxter16} & $\L$ \citep{Baxter16}\tabularnewline
\hline 
\hline 
number of sites & $L$ & $M$ & $M$ & $N$\tabularnewline
reduced coupling & $\K^{\L}$ & $\K^{\M}$ & $H$ & $H'$\tabularnewline
weights & $(z,t)$ & $(t^{*},z^{*})$ & $(t^{*},u^{*})$ & $(u,t)$\tabularnewline
transfer matrix & $\mat{\mathcal{T}}$ & $\mat Q$ & $\widehat{V}_{2}\widehat{V}_{1}$ & $-$\tabularnewline
eigenvalues & $\lambda$ & $\zeta$ & $\lambda$ & $z$\tabularnewline
angles & $\gamma$ & $\ii\varphi$ & $-$ & $-$\tabularnewline
boundary angles & $\theta$ & $\ii\omega$ & $-$ & $-$\tabularnewline
characteristic polynomial~~~ & $1-\ee^{L\gamma-\theta}$ & $1-\ee^{\ii(M\varphi-\omega)}$ & $-$ & \citep[(3.21)]{Baxter16}\tabularnewline
elliptic variables & $(u,\eta)$ & $(\tilde{u},\tilde{\eta})$ & $(r,\nu/2)$ & $(\bar{r},\bar{\nu}/2)$\tabularnewline
elliptic modulus & \multicolumn{2}{c|}{$k$} & \multicolumn{2}{c}{$k^{-1}$}\tabularnewline
$\Im(\text{CP zeroes})$ & $(0,K')$ & $\pm K'/2$ & $-$ & $-$\tabularnewline
\end{tabular}
\end{table}
During the present calculation, it was tried to find a representation
of the partition function $Z$ that is formally symmetric under the
exchange of the two directions, mediated by the transformation $\fliptrans$.
While the symbol function $a(\zeta)$ from (\ref{eq:symbols_a_b})
already fulfills $\fliptrans[a(\zeta)]=1/a(\zeta)$, the transformation
$\fliptrans$ exchanges the system dimensions $L\stackrel{\fliptrans}{\leftrightarrow}M$
and therefore changes the dimensions of the involved matrices, such
that, e.\,g., a representation using $(M{+}L)\times(M{+}L)$ dimensional
matrices might be necessary for a unified description.

The obtained results might possibly be rewritten using elliptic product
identities, as done by Baxter \citep{Baxter16} in the thermodynamic
limit. The ultimate goal would be to use elliptic determinant evaluations
as done successfully by Iorgov \& Lisovyy \citep{IorgovLisovyy11},
which might lead to a closed product representation of the partition
function. 

We expect that our results can be extended to other boundary conditions
by using the corresponding characteristic polynomials, these generalizations
are left for future work. 

From the Toeplitz determinant representation of the partition function,
it is rather straightforward to derive the (anisotropic) scaling limit
$L,M\to\infty$, $T\to\Tc$ at fixed scaling variables $\xM\Def(T/\Tc-1)(M/\xi_{0}^{\M})^{1/\nu}$
and $\rho\Def(L/\xi_{0}^{\L})/(M/\xi_{0}^{\M})$ using Szeg\H{o}'s
theorem \citep{Szego15,Szego20,DeiftItsKrasovsky13}. This task will
be addressed in a forthcoming work. 
\begin{acknowledgments}
The author is grateful to Hendrik Oppenberg, Felix M. Schmidt and
Wolfhard Janke for helpful discussions and inspirations. Discussions
with Jan Büddefeld, Luca Cervellera and Nils Gluth are acknowledged.
This work was partially supported by the Deutsche Forschungsgemeinschaft
through Grant HU~2303/1-1.
\end{acknowledgments}

\appendix

\section{Useful elliptic identities\label{sec:Elliptic-Identities}}

\allowdisplaybreaks

In this appendix we will list identities arising from the elliptic
parametrization. We also simplify expressions given in chapter VI
of \citep{Hucht16a}. Following McCoy \& Wu \citep{McCoyWu14} (who
used $\alpha_{1}^{\pm1},\alpha_{2}^{\pm1}$) and Iorgov \& Lisovyy
\citep{IorgovLisovyy11} (who used $\alpha^{\pm1},\beta^{\pm1}$),
we introduce abbreviations for products and ratios of $z$ and $t$,
however we again utilize Glashier's notation (\ref{eq:Glashier_Graphical})
and define the constants
\begin{equation}
\lan\Def tz,\qquad\las\Def\frac{1}{tz},\qquad\lac\Def\frac{t}{z},\qquad\lad\Def\frac{z}{t},\label{eq:lambda_nscd_Def}
\end{equation}
with $\lan\las=\lac\lad=1$. This leads to the identities
\begin{align}
\Jsn^{2}\eta & =-\lan k^{-1}, & \Jcn^{2}\eta & =\frac{\lan\lan_{,-}}{tt_{-}}=1+\lan k^{-1}, & \Jdn^{2}\eta & =\frac{\lan\lan_{,-}}{zz_{-}}=1+\lan k.\label{eq:sncndn_sq_eta}
\end{align}

Furthermore, We list relations to the primary reduced couplings $\K^{\delta}$
from (\ref{eq:Ising}),\bS
\begin{align}
t & =\exp(-2\K^{\M}), & t_{+} & =\cosh(2\K^{\M}), & t_{-} & =-\sinh(2\K^{\M}),\label{eq:t_of_K^M}\\
z & =\tanh\K^{\L}, & z_{+} & =\coth(2\K^{\L}), & z_{-} & =-\csch(2\K^{\L}),\label{eq:z_of_K^L}
\end{align}
whereas for the dual couplings $z^{*},t^{*}$ the directions $\L$
and $\M$ are to be exchanged,
\begin{align}
z^{*} & =\exp(-2\K^{\L}), & {z^{*}}_{\!+} & =\cosh(2\K^{\L}), & {z^{*}}_{\!-} & =-\sinh(2\K^{\L}),\label{eq:z*_of_K^L}\\
t^{*} & =\tanh\K^{\M}, & {t^{*}}_{\!+} & =\coth(2\K^{\M}), & {t^{*}}_{\!-} & =-\csch(2\K^{\M}).\label{eq:t*_of_K^M}
\end{align}
\eS{}Defining the dual primary reduced couplings $\tilde{\K}^{\delta}$
via
\begin{equation}
\exp(-2\tilde{\K}^{\delta})=\tanh\K^{\delta},\label{eq:tildeK}
\end{equation}
we conclude from (\ref{eqs:tz(am(eta))}) that the following simple
relations hold between $\tilde{\K}^{\delta}$ and $\eta$,
\begin{align}
2\ii\tilde{\K}^{\M} & =\Jam(2\eta), & 2\ii\tilde{\K}^{\L} & =\Jam(2\tilde{\eta}),\label{eq:K_as_am(2eta)}
\end{align}
such that the Jacobi amplitude (\ref{eq:am_Def}) represents a direct
connection between the physical reduced couplings $\K^{\delta}$ and
the parameter $\eta$, and (\ref{eq:eta_from_F}) can be written using
the elliptic Integral of the first kind, cf. (\ref{eq:F(phi,k)}),
\begin{equation}
2\eta=F(2\ii\tilde{\K}^{\M},k)=\ii K'-F(2\ii\tilde{\K}^{\L},k).\label{eq:eta_from_K}
\end{equation}

We now turn to the eigenvalues $\lambda$ and $\zeta$. Defining the
abbreviation
\begin{equation}
Q(u,\eta)\Def\sqrt{\lan-\lambda}=\sqrt{\frac{(k\Jsn^{2}\eta)^{2}-1}{k\Jsn^{2}u-k\Jsn^{2}\eta}},\label{eq:Q_Def}
\end{equation}
we can express the four roots from chapter VI of \citep{Hucht16a}
as meromorphic functions of $u$, eliminating the ambiguous signs
of the square roots, \bS\label{eqs:lambda_xn(u)}
\begin{align}
\sqrt{\lan-\lambda} & =Q(u,\eta)\frac{\Jnn u}{\Jnn\eta}, & \sqrt{\las-\lambda} & =Q(u,\eta)\frac{\Jsn u}{\Jsn\eta},\label{eq:lambda_nn(u)_sn(u)}\\
\sqrt{\lac-\lambda} & =Q(u,\eta)\frac{\Jcn u}{\Jcn\eta}, & \sqrt{\lad-\lambda} & =Q(u,\eta)\frac{\Jdn u}{\Jdn\eta}.\label{eq:lambda_cn(u)_dn(u)}
\end{align}
\eS{}Note that we have used the trivial elliptic function $\Jnn u\Def1$
in order to illustrate the systematics. Using (\ref{eq:flip_u}) we
can derive analog expressions for $\zeta$,\bS\label{eqs:zeta_xn(u)}
\begin{align}
\sqrt{\zeta_{\mathrm{n}}-\zeta} & =Q(\tilde{u},\tilde{\eta})\frac{\Jnn\tilde{u}}{\Jnn\tilde{\eta}}, & \sqrt{\zeta_{\mathrm{s}}-\zeta} & =Q(\tilde{u},\tilde{\eta})\frac{\Jsn\tilde{u}}{\Jsn\tilde{\eta}},\label{eq:zeta_nn(u)_sn(u)}\\
\sqrt{\zeta_{\mathrm{c}}-\zeta} & =Q(\tilde{u},\tilde{\eta})\frac{\Jcn\tilde{u}}{\Jcn\tilde{\eta}}, & \sqrt{\zeta_{\mathrm{d}}-\zeta} & =Q(\tilde{u},\tilde{\eta})\frac{\Jdn\tilde{u}}{\Jdn\tilde{\eta}},\label{eq:zeta_cn(u)_dn(u)}
\end{align}
\eS{}where we defined 
\begin{equation}
\zeta_{\mathrm{n}}\Def z^{*}t^{*},\qquad\zeta_{\mathrm{s}}\Def\frac{1}{z^{*}t^{*}},\qquad\zeta_{\mathrm{c}}\Def\frac{z^{*}}{t^{*}},\qquad\zeta_{\mathrm{d}}\Def\frac{t^{*}}{z^{*}},\label{eq:zeta_nscd_Def}
\end{equation}
in analogy to (\ref{eq:lambda_nscd_Def}). From (\ref{eq:lambda_nn(u)_sn(u)})
and (\ref{eq:zeta_nn(u)_sn(u)}) we further derive \bS{}\label{eqs:lambda(u)-zeta(u)}
\begin{align}
\lambda & =\frac{1-k^{2}\Jsn^{2}u\Jsn^{2}\eta}{k(\Jsn^{2}u-\Jsn^{2}\eta)}, & \zeta & =\frac{1-k^{2}\Jsn^{2}\tilde{u}\Jsn^{2}\tilde{\eta}}{k(\Jsn^{2}\tilde{u}-\Jsn^{2}\tilde{\eta})},\label{eq:lambda(u)-zeta(u)}
\end{align}
which can be expressed using (\ref{eq:dual}),
\begin{align}
\lambda^{*} & =-\frac{(k\Jsn^{2}u)^{*}}{(k\Jsn^{2}\eta)^{*}}, & \zeta^{*} & =-\frac{(k\Jsn^{2}\tilde{u})^{*}}{(k\Jsn^{2}\tilde{\eta})^{*}}.\label{eq:lambda(u)-zeta(u)-dual}
\end{align}
\eS{}Inserting the elliptic expressions (\ref{eqs:lambda_xn(u)})
into \cI{46} we find\bS
\begin{alignat}{2}
\sin\frac{\varphi}{2} & =-\frac{\sqrt{\lac-\lambda}\sqrt{\lad-\lambda}}{2\sqrt{\lambda t_{-}z_{-}}} &  & =-\frac{Q^{2}(u,\eta)}{2\sqrt{\lambda t_{-}z_{-}}}\frac{\Jcn u}{\Jcn\eta}\frac{\Jdn u}{\Jdn\eta},\label{eq:id1-sin}\\
\cos\frac{\varphi}{2} & =\frac{\sqrt{\lan-\lambda}\sqrt{\las-\lambda}}{2\ii\sqrt{\lambda t_{-}z_{-}}} &  & =\frac{Q^{2}(u,\eta)}{2\ii\sqrt{\lambda t_{-}z_{-}}}\frac{\Jnn u}{\Jnn\eta}\frac{\Jsn u}{\Jsn\eta},\label{eq:id1-cos}\\
\tan\frac{\varphi}{2} & =\frac{1}{\ii}\frac{\sqrt{\lac-\lambda}\sqrt{\lad-\lambda}}{\sqrt{\lan-\lambda}\sqrt{\las-\lambda}} &  & =\frac{1}{\ii}\frac{\Jnn\eta}{\Jnn u}\frac{\Jsn\eta}{\Jsn u}\frac{\Jcn u}{\Jcn\eta}\frac{\Jdn u}{\Jdn\eta},\label{eq:id1-tan}
\end{alignat}
\eS{}while \cI{47} at the eigenvalues $\lambda_{\mu}$ become \bS
\begin{alignat}{2}
\pm\sin\frac{M\varphi_{\mu}}{2} & =\sqrt{t}\frac{\sqrt{\las-\lambda_{\mu}}\sqrt{\lad-\lambda_{\mu}}}{2\sqrt{\lambda_{\mu}t_{-}\lambda_{\mu,-}}} &  & =\frac{\sqrt{t}}{2}\frac{Q^{2}(u_{\mu},\eta)}{\sqrt{\lambda_{\mu}t_{-}\lambda_{\mu,-}}}\frac{\Jsn u_{\mu}}{\Jsn\eta}\frac{\Jdn u_{\mu}}{\Jdn\eta},\label{eq:idM-sin}\\
\pm\cos\frac{M\varphi_{\mu}}{2} & =\frac{1}{\sqrt{t}}\frac{\sqrt{\lan-\lambda_{\mu}}\sqrt{\lac-\lambda_{\mu}}}{2\ii\sqrt{\lambda_{\mu}t_{-}\lambda_{\mu,-}}} &  & =\frac{1}{2\ii\sqrt{t}}\frac{Q^{2}(u_{\mu},\eta)}{\sqrt{\lambda_{\mu}t_{-}\lambda_{\mu,-}}}\frac{\Jnn u_{\mu}}{\Jnn\eta}\frac{\Jcn u_{\mu}}{\Jcn\eta},\label{eq:idM-cos}\\
\tan\frac{M\varphi_{\mu}}{2} & =\ii t\frac{\sqrt{\las-\lambda_{\mu}}\sqrt{\lad-\lambda_{\mu}}}{\sqrt{\lan-\lambda_{\mu}}\sqrt{\lac-\lambda_{\mu}}} &  & =\frac{\Jsn u_{\mu}\Jdn u_{\mu}}{\Jnn u_{\mu}\Jcn u_{\mu}}.\label{eq:idM-tan}
\end{alignat}
\eS{}We additionally list the identities\bS
\begin{align}
t_{-}z_{-}\ii\sin\varphi & =-\frac{1}{2\lambda}\sqrt{\lan-\lambda}\sqrt{\las-\lambda}\sqrt{\lac-\lambda}\sqrt{\lad-\lambda}\\
 & =\sqrt{\las_{,+}-\lambda_{+}}\sqrt{\lad_{,+}-\lambda_{+}}\\
 & =-\frac{Q^{4}(u,\eta)}{2\lambda}\frac{\Jnn u}{\Jnn\eta}\frac{\Jsn u}{\Jsn\eta}\frac{\Jcn u}{\Jcn\eta}\frac{\Jdn u}{\Jdn\eta}.
\end{align}
 as well as
\begin{align}
\sqrt{2t_{-}z_{-}}\,\ii\sin\frac{\varphi}{2} & =\sqrt{\lad_{,+}-\lambda_{+}}, & \sqrt{2t_{-}z_{-}}\cos\frac{\varphi}{2} & =\sqrt{\las_{,+}-\lambda_{+}}.
\end{align}
\eS{}

Using the addition theorem \Citep[(2.4.22)]{Lawden89}
\begin{equation}
k\Jsn u\,\Jsn v=k\frac{\Jcn(u-v)-\Jcn(u+v)}{\Jdn(u-v)+\Jdn(u+v)}=\frac{1}{k}\frac{\Jdn(u-v)-\Jdn(u+v)}{\Jcn(u-v)+\Jcn(u+v)}\label{eq:Lawden_2.4.22}
\end{equation}
 we derive the important identities\bS\label{eqs:lambda_with_2u}
\begin{align}
\lambda=\ee^{\gamma} & =-k\frac{\Jcn(2u)+\Jcn(2\eta)}{\Jdn(2u)-\Jdn(2\eta)}=-\frac{1}{k}\frac{\Jdn(2u)+\Jdn(2\eta)}{\Jcn(2u)-\Jcn(2\eta)},\label{eq:lambda_with_2u}\\
\lambda_{+}=\cosh\gamma & =-k\frac{\Jcn(2u)\Jdn(2u)+\Jcn(2\eta)\Jdn(2\eta)}{\Jdn^{2}(2u)-\Jdn^{2}(2\eta)},\label{eq:lambda+_with_2u}\\
\lambda_{-}=\sinh\gamma & =-k\frac{\Jcn(2u)\Jdn(2\eta)+\Jcn(2\eta)\Jdn(2u)}{\Jdn^{2}(2u)-\Jdn^{2}(2\eta)},\label{eq:lambda-_with_2u}
\end{align}
\eS{}and, by the swap transformation (\ref{eq:flip_u}), \bS\label{eqs:zeta_with_2u}
\begin{align}
\zeta=\ee^{\ii\varphi} & =-\frac{\Jds(2u)+\Jds(2\eta)}{\Jcs(2u)-\Jcs(2\eta)}=-\frac{\Jcs(2u)+\Jcs(2\eta)}{\Jds(2u)-\Jds(2\eta)},\label{eq:zeta_with_2u}\\
\zeta_{+}=\cos\varphi & =-\frac{\Jds(2u)\Jcs(2u)+\Jds(2\eta)\Jcs(2\eta)}{\Jcs^{2}(2u)-\Jcs^{2}(2\eta)},\label{eq:zeta+_with_2u}\\
\zeta_{-}=\ii\sin\varphi & =-\frac{\Jds(2u)\Jcs(2\eta)+\Jds(2\eta)\Jcs(2u)}{\Jcs^{2}(2u)-\Jcs^{2}(2\eta)},\label{eq:zeta-_with_2u}
\end{align}
\eS{}which express the eigenvalues $\lambda$ and $\zeta$ as functions
of $2u$. 

We now turn to derivatives. From (\ref{eq:sn(u_pm_eta)_Def}) and
\begin{equation}
\frac{\partial}{\partial u}\log\Jsn(u\pm\eta)=\frac{\Jcn(u\pm\eta)\Jdn(u\pm\eta)}{\Jsn(u\pm\eta)}=k\lambda_{-}[\Jsn(2u)\mp\Jsn(2\eta)]\label{eq:dlogsn_du}
\end{equation}
we see that the derivatives of $\varphi$ and $\gamma$ w.r.t.~$u$
become \bS
\begin{alignat}{3}
\frac{1}{2}\frac{\partial\varphi}{\partial u} & =\,\,\ii k\Jsn(2\eta)\,\lambda_{-} &  & =\frac{1}{z_{-}}\sinh\gamma &  & =-\frac{\sin\varphi}{\sin\omega},\label{eq:dphi_du}\\
\frac{1}{2}\frac{\partial\gamma}{\partial u} & =-k\Jsn(2u)\,\lambda_{-} &  & =t_{-}\sin\varphi &  & =\ii\frac{\sinh\gamma}{\sinh\theta},\label{eq:dgamma_du}
\end{alignat}
\eS{}from which other identities, such as 
\begin{equation}
\frac{\partial\gamma}{\partial\varphi}=-t_{-}\sin\omega=t_{-}z_{-}\frac{\sin\varphi}{\sinh\gamma},\qquad\frac{\partial\chi}{\partial u}=\frac{\chi^{2}-4}{\sin\omega},\label{eq:dgamma_dphi-dchi_du}
\end{equation}
are easily calculated.

In the ordered phase where $k>1$, the angle $\varphi_{1}$ becomes
complex \citep[chap.~6]{Hucht16a}, leading to a complex value of
$u_{1}$. The correct mapping from the eigenvalues $\lambda_{\mu}$
to the elliptic variable $u_{\mu}$, respecting this behavior and
being valid at arbitrary temperatures, can be expressed using the
inverse Jacobi dn, see (\ref{eqs:lambda_xn(u)}),
\begin{equation}
u_{\mu}=\Jdn^{(-1)}\negmedspace\left[\Jdn\eta\,\frac{\sqrt{\lad-\lambda_{\mu}}}{\sqrt{\lan-\lambda_{\mu}}}\right].\label{eq:u_mu_Def}
\end{equation}
While it is tempting to utilize the simpler relation (\ref{eq:omega_Def})
\begin{equation}
u_{\mu}=\frac{1}{2}F(M\varphi_{\mu},k),\label{eq:u_mu_alt}
\end{equation}
it won't give correct results for even $\mu$ and below $\Tc$, because
the elliptic integral $F$ does not have the correct branch cut positions
for these cases.

\section{A block Hankel matrix identity \label{sec:Block-Hankel-matrix}}

Let $\mat{\mathcal{V}}_{\boldsymbol{g},\boldsymbol{x}}$ be the generalized
$1\times B$ block Vandermonde matrix with $M\times N$ blocks
\begin{equation}
\mat{\mathcal{V}}_{\boldsymbol{g},\boldsymbol{x}}\Def\left[\vphantom{A_{2}^{2}}g_{\mu}^{b}\,x_{\mu}^{n}\right]_{b=0;\,\,\mu=1,\,n=0}^{B-1\,M\hspace{2ex}N-1}=\left[\vphantom{A_{2}^{2}}\mat G^{b}\,\mat V_{\!\boldsymbol{x}}\right]_{b=0}^{B-1},\label{eq:V_g,x}
\end{equation}
 with 
\begin{equation}
\mat V_{\!\boldsymbol{x}}\Def\left[\vphantom{A_{2}^{2}}x_{\mu}^{n}\right]_{\mu=1,\,n=0}^{M\hspace{2ex}N-1},\qquad\mat G\Def\left[\vphantom{A_{2}^{2}}\delta_{\mu\nu}\,g_{\mu}\right]_{\mu,\nu=1}^{M}.\label{eq:V_x-1}
\end{equation}
As an example, for $M=4$, $B=3$ and $N=3$ we have \begingroup\renewcommand*{\arraystretch}{1.0}
\begin{equation}
\mat{\mathcal{V}}_{\boldsymbol{g},\boldsymbol{x}}=\setlength{\arraycolsep}{4pt}\left[\begin{array}{ccc|ccc|ccc}
1 & x_{1} & x_{1}^{2} & g_{1} & g_{1}x_{1} & g_{1}x_{1}^{2} & g_{1}^{2} & g_{1}^{2}x_{1} & g_{1}^{2}x_{1}^{2}\\
1 & x_{2} & x_{2}^{2} & g_{2} & g_{2}x_{2} & g_{2}x_{2}^{2} & g_{2}^{2} & g_{2}^{2}x_{2} & g_{2}^{2}x_{2}^{2}\\
1 & x_{3} & x_{3}^{2} & g_{3} & g_{3}x_{3} & g_{3}x_{3}^{2} & g_{3}^{2} & g_{3}^{2}x_{3} & g_{3}^{2}x_{3}^{2}\\
1 & x_{4} & x_{4}^{2} & g_{4} & g_{4}x_{4} & g_{4}x_{4}^{2} & g_{4}^{2} & g_{4}^{2}x_{4} & g_{4}^{2}x_{4}^{2}
\end{array}\right].\label{V_g,x-4}
\end{equation}
\endgroup{}Furthermore, let $\mat D$ be an arbitrary $M\times M$
diagonal matrix. Then, the $B\times B$ block Hankel matrix with $N\times N$
blocks
\begin{equation}
\mat{\mathcal{H}}_{\boldsymbol{g},\boldsymbol{x}}\Def\left[\sum_{\mu=1}^{M}d_{\mu}g_{\mu}^{a+b}\,x_{\mu}^{m+n}\right]_{a,b=0;\,m,n=0}^{B-1\;\;\;N-1}=\left[\T{\mat V}_{\!\boldsymbol{x}}\,\mat D\,\mat G^{a+b}\,\mat V_{\!\boldsymbol{x}}\right]_{a,b=0}^{B-1}\label{eq:H_g,x}
\end{equation}
trivially fulfills the identity
\begin{equation}
\mat{\mathcal{H}}_{\boldsymbol{g},\boldsymbol{x}}=\T{\mat{\mathcal{V}}}_{\!\boldsymbol{g},\boldsymbol{x}}\mat D\mat{\mathcal{V}}_{\!\boldsymbol{g},\boldsymbol{x}}.\label{eq:H_g,x-id}
\end{equation}
Note that the upper left block $(\mat{\mathcal{H}}_{\boldsymbol{g},\boldsymbol{x}})_{0,0}=\T{\mat V}_{\!\boldsymbol{x}}\,\mat D\,\mat V_{\!\boldsymbol{x}}$
is free of $g_{\mu}$ and is therefore a usual Vandermonde product
in $x_{\mu}$. Consequently, if the matrix $\mat D$ is set to the
diagonal matrix $\mat D_{\boldsymbol{x}}$ with the reciprocal first
derivatives of the characteristic polynomial $\P x(x)$ from (\ref{eq:D_x,H_x}),
\begin{equation}
\mat D_{\boldsymbol{x}}\Def\left[\frac{\delta_{\mu\nu}}{\P x'(x_{\mu})}\right]_{\mu,\nu=1}^{M},\label{eq:D_x}
\end{equation}
and if additionally $M\geq2N$, then $(\mat H_{\boldsymbol{g},\boldsymbol{x}})_{0,0}$
vanishes identically. For $B=2$ and $N=M/2$, this leads to equation
(\ref{eq:generalized_V_H_Identity}).

\section{More characteristic polynomials}

In \citep{Hucht16b} we used the finite-size scaling limit of $\P{\lambda_{+}}$
to locate the zeroes in the complex plane and to perform the corresponding
Cauchy integrals. We had to distinguish between even and odd zeroes
and defined an alternating counting polynomial from $\P{\lambda_{+}}$.
Now we will demonstrate that it is much easier to analyze the complex
structure of the system by using the CP of the transfer matrix $\mat T$
instead of $\mat T_{+}$. While it is possible but cumbersome to derive
the CP of $\mat T$, with eigenvalues $\lambda_{\mu}$, 
\begin{equation}
\P{\lambda}(\lambda)\Def\det(\lambda\mat 1-\mat T)=\prod_{\mu=1}^{M}(\lambda-\lambda_{\mu})\label{eq:CP_lambda_Def}
\end{equation}
from scratch analogously to (\ref{eq:CP+_Def}), cf.~\citep{Hucht16a},
we instead proceed in a much simpler way and derive it directly from
(\ref{eq:CP+(u)}): Using (\ref{eq:apm}), we first factorize the
right hand side of (\ref{eq:CP+_Def}),
\begin{equation}
\P{\lambda_{+}}(\lambda_{+})=\prod_{\mu=1}^{M}(\lambda_{+}-\lambda_{\mu,+})=\prod_{\mu=1}^{M}\frac{(\lambda-\lambda_{\mu})(\lambda^{-1}-\lambda_{\mu})}{2(0-\lambda_{\mu})}=\frac{1}{2^{M}t}\P{\lambda}(\lambda)\P{\lambda}(\lambda^{-1}),\label{eq:CP+_Factor}
\end{equation}
as $\lambda_{\mu,+}=\lambda_{+.\mu}$ and, cf.~\cI{35},
\begin{equation}
\P{\lambda}(0)=\det(-\mat T)=\det\mat T=t.\label{eq:P(0)}
\end{equation}
Employing the trigonometric factorization identity
\begin{equation}
\frac{\sin\left(M\varphi-\omega\right)}{\sin(-\omega)}=\frac{\sin\left({\textstyle \frac{1}{2}}[M\varphi-\omega]\right)}{\sin\left(-{\textstyle \frac{1}{2}}\omega\right)}\frac{\cos\left({\textstyle \frac{1}{2}}[M\varphi-\omega]\right)}{\cos\left(-{\textstyle \frac{1}{2}}\omega\right)}\label{eq:trig_id}
\end{equation}
as well as the identities 
\begin{equation}
\lambda(\ui)\lambda(u)=1,\quad\varphi(\ui)=-\varphi(u),\quad\omega(\ui)+\omega(u)=\pi,\quad\cot[{\textstyle \frac{1}{2}}\omega(\ui)]=\tan[{\textstyle \frac{1}{2}}\omega(u)],\label{eq:inv_trans}
\end{equation}
with inversion transform $u\mapsto\ui\Def u+\ii K'$, we see that
$\P{\lambda}(\lambda)$ and $\P{\lambda}(\lambda^{-1})$ are given
by the remarkably simple formulas \bS\label{eq:CP(u)}
\begin{align}
\P{\lambda}(\lambda) & =(1-t^{*})\left(-t_{-}z_{-}\lambda\right)^{\frac{M}{2}}\left[\cos\left({\textstyle \frac{1}{2}}M\varphi\right)-\cot\left({\textstyle \frac{1}{2}}\omega\right)\sin\left({\textstyle \frac{1}{2}}M\varphi\right)\right]\label{eq:CP(u)_1}\\
 & =(1-t^{*})\left(-t_{-}z_{-}\lambda\right)^{\frac{M}{2}}\frac{\sin\left({\textstyle \frac{1}{2}}[M\varphi-\omega]\right)}{\sin\left(-{\textstyle \frac{1}{2}}\omega\right)}\label{eq:CP(u)_2}\\
 & =(1-t^{*})\left(-\frac{t_{-}z_{-}\lambda}{\zeta}\right)^{\negmedspace\frac{M}{2}}\frac{1-\ee^{\ii(M\varphi-\omega)}}{1-\ee^{-\ii\omega}},\label{eq:CP(u)_3}\\
\P{\lambda}(\lambda^{-1}) & =(1-t^{*})\left(-t_{-}z_{-}\lambda^{-1}\right)^{\frac{M}{2}}\left[\cos\left({\textstyle \frac{1}{2}}M\varphi\right)+\tan\left({\textstyle \frac{1}{2}}\omega\right)\sin\left({\textstyle \frac{1}{2}}M\varphi\right)\right]\label{eq:CP(ui)_1}\\
 & =(1-t^{*})\left(-t_{-}z_{-}\lambda^{-1}\right)^{\frac{M}{2}}\frac{\cos\left({\textstyle \frac{1}{2}}[M\varphi-\omega]\right)}{\cos\left(-{\textstyle \frac{1}{2}}\omega\right)}\label{eq:CP(ui)_2}\\
 & =(1-t^{*})\left(-\frac{t_{-}z_{-}}{\lambda\zeta}\right)^{\negmedspace\frac{M}{2}}\frac{1+\ee^{\ii(M\varphi-\omega)}}{1+\ee^{-\ii\omega}},\label{eq:CP(ui)_3}
\end{align}
\eS{}The additional factor $(-\lambda)^{M/2}$ in $\P{\lambda}(\lambda)$
follows from (\ref{eq:P(0)}) in the known limits $\lambda\to\{0,\infty\}$,
see (\ref{eqs:counter_poles}). Utilizing a factorization similar
to (\ref{eq:CP+_Factor}), the CP of $\mat T_{-}$ can also be derived,
\begin{equation}
\P{\lambda_{-}}(\lambda_{-})=\prod_{\mu=1}^{M}(\lambda_{-}-\lambda_{\mu,-})=\prod_{\mu=1}^{M}\frac{(\lambda-\lambda_{\mu})(-\lambda^{-1}-\lambda_{\mu})}{2(0-\lambda_{\mu})}=\frac{1}{2^{M}t}\P{\lambda}(\lambda)\P{\lambda}(-\lambda^{-1}).\label{eq:CP-}
\end{equation}
Finally, from (\ref{eq:CP_chi_factor}) and Liouville's theorem \citep[15.3]{Baxter82}
we derived the CPs 
\begin{align}
\P{\zeta}(\zeta) & =(1-t^{*})\frac{1-\ee^{\ii(M\varphi-\omega)}}{1-\ee^{-\ii\omega}}\prod_{\mu=1}^{M}\frac{\Jsn(\eta+u_{\mu})}{\Jsn(u+u_{\mu})},\label{eq:CP_zeta(zeta)}\\
\P{\zeta}(\zeta^{-1}) & =(1-t^{*})\frac{1+\ee^{-\ii(M\varphi-\omega)}}{1+\ee^{\ii\omega}}\prod_{\mu=1}^{M}k\Jsn(\eta+u_{\mu})\Jsn(u+u_{\mu}).\label{eq:CP_zeta(1/zeta)}
\end{align}

\section{Some product identities}

Using the CPs (\ref{eq:CP+(u)}), (\ref{eq:CP(u)}), (\ref{eq:CP-}),
we have the following identities (remember that $M$ is even): the
determinants are given by \bS
\begin{align}
\det\mat T_{\hphantom{-}} & =\prod_{\mu=1}^{M}\lambda_{\mu\hphantom{,-}}=\P{\lambda}(0)_{\hphantom{-}}=t\,,\label{eq:det_T}\\
\det\mat T_{-} & =\prod_{\mu=1}^{M}\lambda_{\mu,-}=\P{\lambda_{-}}(0)=(1-{t^{*}}^{2})\left(\frac{\ii z_{-}}{1-{t^{*}}^{2}}\right)^{\negmedspace M}\negmedspace,\label{eq:det_T-}\\
\det\mat T_{+} & =\prod_{\mu=1}^{M}\lambda_{\mu,+}=\P{\lambda_{+}}(0)=(1-{t^{*}}^{2})\left(\frac{t_{-}z_{-}}{2}\right)^{\negmedspace M}\frac{\sin(M\varphi_{\lambda_{+}\to0}-\omega_{\lambda_{+}\to0})}{\sin(-\omega_{\lambda_{+}\to0})},\label{eq:det_T+}
\end{align}
\eS{}with (\ref{eqs:counter_poles}) and
\begin{equation}
u_{\lambda_{+}\to0}=\Jsn^{(-1)}\sqrt{\frac{(\ii\lan)^{*}}{\ii k}},\qquad u_{\lambda_{-}\to0}=K+\tfrac{1}{2}\ii K'.\label{eqs:u_lambda_pm=00003D0}
\end{equation}
 Furthermore, we have the following product identities for the Jacobi
elliptic functions\bS
\begin{align}
\prod_{\mu=1}^{M}\sqrt{-tz}\,\frac{\Jsn u_{\mu}}{\Jsn\eta} & =\sqrt{1-M\las_{,-}z_{-}^{-1}}\,,\label{eq:prod_sn}\\
\prod_{\mu=1}^{M}\sqrt{\ii z}\,\frac{\Jcn u_{\mu}}{\Jcn\eta} & =1\,,\label{eq:prod_cn}\\
\prod_{\mu=1}^{M}\sqrt{\ii t}\,\frac{\Jdn u_{\mu}}{\Jdn\eta} & =\sqrt{1+M\lad_{,-}z_{-}^{-1}}\,.\label{eq:prod_dn}
\end{align}
\eS{}For products over $\lambda_{\mu}$ we find the identities \bS
\begin{align}
\prod_{\mu=1}^{M}(\lan-\lambda_{\mu}) & =\P{\lambda}(\lan)=(1-t^{*})(t_{-}z_{-}\lan)^{\frac{M}{2}},\label{eq:prod_lambda_n}\\
\prod_{\mu=1}^{M}(\lac-\lambda_{\mu}) & =\P{\lambda}(\lac)=(1-t^{*})(-t_{-}z_{-}\lac)^{\frac{M}{2}},\label{eq:prod_lambda_c}\\
\prod_{\mu=1}^{M}(\las-\lambda_{\mu}) & =\P{\lambda}(\las)=(1-t^{*})(t_{-}z_{-}\las)^{\frac{M}{2}}\left(1-M\las_{,-}z_{-}^{-1}\right),\label{eq:prod_lambda_s}\\
\prod_{\mu=1}^{M}(\lad-\lambda_{\mu}) & =\P{\lambda}(\lad)=(1-t^{*})(-t_{-}z_{-}\lad)^{\frac{M}{2}}\left(1+M\lad_{,-}z_{-}^{-1}\right),\label{eq:prod_lambda_d}
\end{align}
\eS{}and for products over $\zeta_{\mu}$ we derive\bS
\begin{align}
\prod_{\mu=1}^{M}\sin\frac{\varphi_{\mu}}{2} & =\frac{1-t^{*}}{(2\ii)^{M}\sqrt{t}}\sqrt{1+M\lad_{,-}z_{-}^{-1}}\,,\label{eq:prod_sin}\\
\prod_{\mu=1}^{M}\cos\frac{\varphi_{\mu}}{2} & =\frac{1-t^{*}}{2^{M}\sqrt{t}}\sqrt{1-M\las_{,-}z_{-}^{-1}}\,,\label{eq:prod_cos}\\
\prod_{\mu=1}^{M}\tan\frac{\varphi_{\mu}}{2} & =(-1)^{M/2}\,\frac{\sqrt{1+M\lad_{,-}z_{-}^{-1}}}{\sqrt{1-M\las_{,-}z_{-}^{-1}}}=\prod_{\mu=1}^{M}\ee^{-\theta_{\mu}}=\prod_{\mu=1}^{M}\ee^{-\psi_{\mu}}\,.\label{eq:prod_tan}
\end{align}
\eS{}Finally, from the factorization
\begin{equation}
(t_{-}z_{-}\ii\sin\varphi)^{2}=(t_{+}z_{+}-\lambda_{+})^{2}-t_{-}^{2}z_{-}^{2}=(\las_{,+}-\lambda_{+})(\lad_{,+}-\lambda_{+}),\label{eq:sin_factorization}
\end{equation}
as $\las_{,+}=t_{+}z_{+}+t_{-}z_{-}$ and $\lad_{,+}=t_{+}z_{+}-t_{-}z_{-}$,
we deduce the closed form expression
\begin{align}
\prod_{\mu=1}^{M}(t_{-}z_{-}\ii\sin\varphi_{\mu})^{2} & =\prod_{\mu=1}^{M}(\las_{,+}-\lambda_{\mu,+})(\lad_{,+}-\lambda_{\mu,+})=\P{\lambda_{+}}(\las_{,+})\P{\lambda_{+}}(\lad_{,+})\nonumber \\
 & =(1-t^{*2})^{2}\left(\frac{t_{-}z_{-}}{2}\right)^{\negmedspace2M}\left(1-M\frac{\las_{,-}}{z_{-}}\right)\left(1+M\frac{\lad_{,-}}{z_{-}}\right).\label{eq:prod_sin_phi}
\end{align}

\bibliographystyle{unsrt}
\bibliography{Physik}

\begin{thebibliography}{10}

\bibitem{Ising25}
E.~Ising.
\newblock {Beitrag zur Theorie des Ferromagnetismus}.
\newblock {\em Z.~Phys.}, 31:253, 1925.

\bibitem{Onsager44}
L.~Onsager.
\newblock Crystal statistics. {I.} {A} two-dimensional model with an
  order-disorder transition.
\newblock {\em Phys. Rev.}, 65:117, 1944.

\bibitem{Kaufman49}
B.~Kaufman.
\newblock Crystal statistics. {II}. {Partition} function evaluated by spinor
  analysis.
\newblock {\em Phys. Rev.}, 76(8):1232--1243, Oct 1949.

\bibitem{McCoyWu73}
B.~M. McCoy and T.~T. Wu.
\newblock {\em The Two-Dimensional {Ising} Model}.
\newblock Harvard University Press, Cambridge, 1973.

\bibitem{Baxter82}
R.~J. Baxter.
\newblock {\em Exactly Solved Models in Statistical Mechanics}.
\newblock Academic Press, London, 1982.

\bibitem{Abraham86}
D.~B. Abraham.
\newblock Surface structures and phase transitions--exact results.
\newblock In C.~Domb and J.L. Lebowitz, editors, {\em Phase Transition and
  Critical Phenomena, Volume 10}, pages 1--74. Academic Press, London, 1986.

\bibitem{Baxter16}
R.~J. Baxter.
\newblock The bulk, surface and corner free energies of the square lattice
  {Ising} model.
\newblock {\em J.~Phys.~A: Math. Theor.}, 50(1):014001, 2017.
\newblock {arXiv:1606.02029}.

\bibitem{Baxter20}
R.~J. Baxter.
\newblock The bulk, surface and corner free energies of the anisotropic
  triangular {Ising} model.
\newblock {\em Proc. Roy. Soc. London~A}, 476(2234):20190713, 2020.

\bibitem{Hucht16a}
Alfred Hucht.
\newblock The square lattice {Ising} model on the rectangle {I}: finite
  systems.
\newblock {\em J.~Phys.~A: Math. Theor.}, 50(6):065201, Jan 2017.
\newblock {arXiv:1609.01963}, erratum \cite{Hucht16ae}.

\bibitem{Hucht16ae}
Alfred Hucht.
\newblock Erratum: The square lattice {Ising} model on the rectangle {I}:
  finite systems.
\newblock {\em J.~Phys.~A: Math. Theor.}, 51(31):319601, Jun 2018.

\bibitem{Hucht16b}
Alfred Hucht.
\newblock The square lattice {Ising} model on the rectangle {II}: finite-size
  scaling limit.
\newblock {\em J.~Phys.~A: Math. Theor.}, 50(26):265205, Jun 2017.
\newblock {arXiv:1701.08722}.

\bibitem{VernierJacobsen12}
Eric Vernier and Jesper~Lykke Jacobsen.
\newblock Corner free energies and boundary effects for {Ising}, {Potts} and
  fully-packed loop models on the square and triangular lattices.
\newblock {\em J.~Phys.~A: Math. Theor.}, 45:045003, 2012.
\newblock arXiv:1110.2158.

\bibitem{Kasteleyn61}
P.~W. Kasteleyn.
\newblock The statistics of dimers on a lattice: {I}. {The} number of dimer
  arrangements on a quadratic lattice.
\newblock {\em Physica}, 27(12):1209 -- 1225, 1961.

\bibitem{Kasteleyn63}
P.~W. Kasteleyn.
\newblock Dimer statistics and phase transitions.
\newblock {\em J.~Math. Phys.}, 4:287, 1963.

\bibitem{Fisher66}
Michael~E. Fisher.
\newblock On the dimer solution of planar {I}sing models.
\newblock {\em Journal of Mathematical Physics}, 7(10):1776--1781, Oct 1966.

\bibitem{McCoyWu14}
B.~M. McCoy and T.~T. Wu.
\newblock {\em The Two-Dimensional {Ising} Model}.
\newblock Dover Books on Physics. Dover Publication, Inc., Mineola, New York,
  2014.

\bibitem{Molinari08}
Luca~G. Molinari.
\newblock Determinants of block tridiagonal matrices.
\newblock {\em Linear Algebra Appl.}, 429:2221, 2008.
\newblock {arXiv:0712.0681}.

\bibitem{FisherdeGennes78}
M.~E. Fisher and P.-G. de~Gennes.
\newblock Ph\'enom\`enes aux parois dans un m\'elange binaire critique.
\newblock {\em C. R. Acad. Sci. Paris, Ser. B}, 287:207, 1978.

\bibitem{FisherAu-Yang80}
Michael~E. Fisher and Helen Au-Yang.
\newblock Critical wall perturbations and a local free energy functional.
\newblock {\em Physica A: Statistical Mechanics and its Applications},
  101(1):255--264, Apr 1980.

\bibitem{CasimirPolder48}
H.~B.~G. Casimir and D.~Polder.
\newblock The influence of retardation on the {London}-van der {Waals} forces.
\newblock {\em Phys. Rev.}, 73:360--372, Feb 1948.

\bibitem{Casimir48}
H.~B.~G. Casimir.
\newblock On the attraction between two perfectly conducting plates.
\newblock {\em Proc. K. Ned. Akad. Wet.}, 51:793, 1948.

\bibitem{HuchtGruenebergSchmidt11}
Alfred Hucht, Daniel Gr\"uneberg, and Felix~M. Schmidt.
\newblock Aspect-ratio dependence of thermodynamic {Casimir} forces.
\newblock {\em Phys. Rev.~E}, 83:051101, Mar 2011.

\bibitem{Kadanoff66}
L.~P. Kadanoff.
\newblock Scaling laws for {Ising} models near {$T_c$}.
\newblock {\em Physics}, 2:263, 1966.

\bibitem{Diehl97a}
H.~W. Diehl.
\newblock The theory of boundary critical phenomena.
\newblock {\em Int. J.~Mod. Phys.~B}, 11(30):3503--3523, 1997.

\bibitem{EvansStecki94}
R.~Evans and J.~Stecki.
\newblock Solvation force in two-dimensional {Ising} strips.
\newblock {\em Phys. Rev. B}, 49:8842--8851, Apr 1994.

\bibitem{Au-YangFisher80}
Helen Au-Yang and Michael~E. Fisher.
\newblock Wall effects in critical systems: {Scaling} in {Ising} model strips.
\newblock {\em Phys. Rev.~B}, 21:3956, 1980.

\bibitem{BrankovDantchevTonchev00}
J.~G. Brankov, D.~M. Dantchev, and N.~S. Tonchev.
\newblock {\em Theory of Critical Phenomena in Finite-Size Systems -- Scaling
  and Quantum Effects}.
\newblock World Scientific, Singapore, 2000.

\bibitem{Gambassi09a}
Andrea Gambassi.
\newblock The {Casimir} effect: From quantum to critical fluctuations.
\newblock {\em Journal of Physics: Conference Series}, 161(1):012037, 2009.

\bibitem{RudnickZandiShackellAbraham10}
Joseph Rudnick, Roya Zandi, Aviva Shackell, and Douglas Abraham.
\newblock Boundary conditions and the critical {Casimir} force on an {Ising}
  model film: Exact results in one and two dimensions.
\newblock {\em Phys. Rev.~E}, 82(4):041118, Oct 2010.

\bibitem{AbrahamMaciolek10}
Douglas~B. Abraham and Anna Macio\l{}ek.
\newblock {Casimir} interactions in {Ising} strips with boundary fields: Exact
  results.
\newblock {\em Phys. Rev. Lett.}, 105:055701, Jul 2010.

\bibitem{AbrahamMaciolek13}
Douglas~B. Abraham and Anna Macio{\l}ek.
\newblock Surface states and the {Casimir} interaction in the {Ising} model.
\newblock {\em {EPL} (Europhysics Letters)}, 101(2):20006, Jan 2013.

\bibitem{Hasenbusch0905}
Martin Hasenbusch.
\newblock The thermodynamic {Casimir} effect in the neighbourhood of the
  lambda-transition: A {Monte} {Carlo} study of an improved three-dimensional
  lattice model.
\newblock {\em J.~Stat. Mech.}, 2009:P07031, 2009.
\newblock arXiv:0905.2096.

\bibitem{Hasenbusch0907}
Martin Hasenbusch.
\newblock Specific heat, internal energy, and the thermodynamic {Casimir} force
  in the neighbourhood of the lambda transition.
\newblock {\em Phys. Rev.~B}, 81:165412, 2010.
\newblock arXiv:0907.2847.

\bibitem{Hasenbusch0908}
Martin Hasenbusch.
\newblock Yet another method to compute the thermodynamic {Casimir} force in
  lattice models.
\newblock {\em Phys. Rev.~E}, 80:061120, 2009.
\newblock arXiv:0908.3582.

\bibitem{Hasenbusch1005}
Martin Hasenbusch.
\newblock Thermodynamic {Casimir} effect for films in the three-dimensional
  {Ising} universality class: Symmetry-breaking boundary conditions.
\newblock {\em Phys. Rev.~B}, 82(10):104425, Sep 2010.
\newblock arXiv:1005.4749.

\bibitem{Hasenbusch1104}
Martin Hasenbusch.
\newblock Thermodynamic {Casimir} force: A {Monte} {Carlo} study of the
  crossover between the ordinary and the normal surface universality class.
\newblock {\em Phys. Rev.~B}, 83:134425, Apr 2011.

\bibitem{Hasenbusch1205}
Martin Hasenbusch.
\newblock Thermodynamic {Casimir} effect: Universality and corrections to
  scaling.
\newblock {\em Phys. Rev.~B}, 85:174421, May 2012.

\bibitem{Hucht07a}
Alfred Hucht.
\newblock Thermodynamic {Casimir} effect in $^{4}${He} films near
  {$T_\lambda$}: {Monte} {Carlo} results.
\newblock {\em Phys. Rev. Lett.}, 99(18):185301, Nov 2007.

\bibitem{MaciolekGambassiDietrich07}
A.~Macio{\l}ek, A.~Gambassi, and S.~Dietrich.
\newblock Critical {Casimir} effect in superfluid wetting films.
\newblock {\em Phys. Rev.~E}, 76:031124, 2007.

\bibitem{VasilyevGambassiMaciolekDietrich07}
O.~Vasilyev, A.~Gambassi, A.~Macio{\l}ek, and S.~Dietrich.
\newblock {Monte} {Carlo} simulation results for critical {Casimir} forces.
\newblock {\em EPL}, 80:60009, 2007.

\bibitem{VasilyevGambassiMaciolekDietrich09}
O.~Vasilyev, A.~Gambassi, A.~Macio{\l}ek, and S.~Dietrich.
\newblock Universal scaling functions of critical {Casimir} forces obtained by
  {Monte} {Carlo} simulations.
\newblock {\em Phys. Rev.~E}, 79(4):041142, 2009.

\bibitem{HobrechtHucht14}
Hendrik Hobrecht and Alfred Hucht.
\newblock Direct simulation of critical {C}asimir forces.
\newblock {\em EPL}, 106(5):56005, Jun 2014.
\newblock {arXiv:1405.4088}.

\bibitem{GarciaChan99url}
R.~Garcia and M.~H.~W. Chan.
\newblock Critical fluctuation-induced thinning of $^4${He} films near the
  superfluid transition.
\newblock {\em Phys. Rev. Lett.}, 83:1187, 1999.

\bibitem{GarciaChan00a}
R.~Garcia and M.~H.~W. Chan.
\newblock Critical {Casimir} effect in dilute $^3${He}-$^4${He} mixture films.
\newblock {\em Physica B}, 280(1):55, 2000.

\bibitem{GarciaChan00b}
R.~Garcia and M.~H.~W. Chan.
\newblock Preliminary measurement of the critical {Casimir} effect near the
  tricritical point in $^3${He}-$^4${He} mixture films.
\newblock {\em J. Low Temp. Phys.}, 121:495, 2000.

\bibitem{GarciaChan02}
R.~Garcia and M.~H.~W. Chan.
\newblock Critical {Casimir} effect near the $^3${He}-$^4${He} tricritical
  point.
\newblock {\em Phys. Rev. Lett.}, 88:086101, 2002.

\bibitem{FukutoYanoPershan05}
M.~Fukuto, Y.~F. Yano, and P.~S. Pershan.
\newblock Critical {Casimir} effect in three-dimensional {Ising} systems:
  Measurements on binary wetting films.
\newblock {\em Phys. Rev. Lett.}, 94:135702, 2005.

\bibitem{GanshinScheidemantelGarciaChan06}
A.~Ganshin, S.~Scheidemantel, R.~Garcia, and M.~H.~W. Chan.
\newblock Critical {Casimir} force in $^4${He} films: Confirmation of
  finite-size scaling.
\newblock {\em Phys. Rev. Lett.}, 97:075301, 2006.

\bibitem{HertleinHeldenGambassiDietrichBechinger08}
C.~Hertlein, L.~Helden, A.~Gambassi, S.~Dietrich, and C.~Bechinger.
\newblock Direct measurement of critical {Casimir} forces.
\newblock {\em Nature}, 451:172, 2008.

\bibitem{GamMacHerNelHelBechDiet09}
A.~Gambassi, A.~Macio\l{}ek, C.~Hertlein, U.~Nellen, L.~Helden, C.~Bechinger,
  and S.~Dietrich.
\newblock Critical {Casimir} effect in classical binary liquid mixtures.
\newblock {\em Phys. Rev. E}, 80(6):061143, Dec 2009.

\bibitem{Polyakov70}
A.~M. Polyakov.
\newblock Conformal symmetry of critical fluctuations.
\newblock {\em JETP Lett.}, 12(12):381, 1970.

\bibitem{Cardy84}
J.~Cardy.
\newblock Conformal invariance and universality in finite-size scaling.
\newblock {\em J.~Phys.~A: Math. Gen.}, 17:L385, 1984.

\bibitem{BurkhardtEisenriegler95}
Theodore~W. Burkhardt and Erich Eisenriegler.
\newblock {Casimir} interaction of spheres in a fluid at the critical point.
\newblock {\em Phys. Rev. Lett.}, 74:3189--3192, Apr 1995.

\bibitem{Cardy2006}
J.~Cardy.
\newblock Boundary conformal field theory.
\newblock In Jean-Pierre Fran\c{c}oise, Gregory~L. Naber, and Tsou~Sheung Tsun,
  editors, {\em Encyclopedia of Mathematical Physics}, pages 333 -- 340.
  Academic Press, Oxford, 2006.

\bibitem{BimonteEmigKardar13}
G.~Bimonte, T.~Emig, and M.~Kardar.
\newblock Conformal field theory of critical {Casimir} interactions in {2D}.
\newblock {\em EPL (Europhysics Letters)}, 104(2):21001, 2013.

\bibitem{FerdinandFisher69}
A.~E. Ferdinand and M.~E. Fisher.
\newblock Bounded and inhomogeneous {Ising} models. {I.} {Specific}-heat
  anomaly of a finite lattice.
\newblock {\em Phys. Rev.}, 185:832, 1969.

\bibitem{LuWu01}
Wentao~T. Lu and F.~Y. Wu.
\newblock {Ising} model on nonorientable surfaces: {Exact} solution for the
  {M{\"o}bius} strip and the {Klein} bottle.
\newblock {\em Phys. Rev.~E}, 63:026107, 2001.

\bibitem{KlebanVassileva91}
P~Kleban and I~Vassileva.
\newblock Free energy of rectangular domains at criticality.
\newblock {\em J.~Phys.~A: Math. Gen.}, 24:3407, 1991.

\bibitem{WuIzmailianGuo12}
Xintian Wu, Nickolay Izmailian, and Wenan Guo.
\newblock Finite-size behavior of the critical {Ising} model on a rectangle
  with free boundaries.
\newblock {\em Phys. Rev. E}, 86:041149, Oct 2012.

\bibitem{SchlesenerHankeDietrich03}
F.~Schlesener, A.~Hanke, and S.~Dietrich.
\newblock Critical {Casimir} forces in colloidal suspensions.
\newblock {\em Journal of Statistical Physics}, 110(3-6):981--1013, 2003.

\bibitem{KHD09}
S.~Kondrat, L.~Harnau, and S.~Dietrich.
\newblock Critical {Casimir} interaction of ellipsoidal colloids with a planar
  wall.
\newblock {\em The Journal of Chemical Physics}, 131(20):204902, 2009.

\bibitem{TrKoGamHarDiet09}
M.~Tr\"{o}ndle, S.~Kondrat, A.~Gambassi, L.~Harnau, and S.~Dietrich.
\newblock Normal and lateral critical {Casimir} forces between colloids and
  patterned substrates.
\newblock {\em EPL}, 88(4):40004, 2009.

\bibitem{GambassiDietrich10}
Andrea Gambassi and S.~Dietrich.
\newblock Colloidal aggregation and critical {Casimir} forces.
\newblock {\em Phys. Rev. Lett.}, 105:059601, Jul 2010.

\bibitem{TrZvGamVogtHarBechDiet11}
Matthias Tr\"{o}ndle, Olga Zvyagolskaya, Andrea Gambassi, Dominik Vogt, Ludger
  Harnau, Clemens Bechinger, and Siegfried Dietrich.
\newblock Trapping colloids near chemical stripes via critical {Casimir}
  forces.
\newblock {\em Molecular Physics}, 109(7-10):1169--1185, 2011.

\bibitem{Hasenbusch13}
Martin Hasenbusch.
\newblock Thermodynamic {Casimir} forces between a sphere and a plate: {Monte}
  {Carlo} simulation of a spin model.
\newblock {\em Phys. Rev. E}, 87:022130, Feb 2013.

\bibitem{LTHD14}
M.~Labbe-Laurent, M.~Trondle, L.~Harnau, and S.~Dietrich.
\newblock Alignment of cylindrical colloids near chemically patterned
  substrates induced by critical {Casimir} torques.
\newblock {\em Soft Matter}, 10:2270--2291, 2014.

\bibitem{HobrechtHucht15a}
Hendrik Hobrecht and Alfred Hucht.
\newblock Many-body critical {C}asimir interactions in colloidal suspensions.
\newblock {\em Phys. Rev.~E}, 92:042315, Oct 2015.

\bibitem{ETBEvRD15}
John~R. Edison, Nikos Tasios, Simone Belli, Robert Evans, Ren\'e van Roij, and
  Marjolein Dijkstra.
\newblock Critical {Casimir} forces and colloidal phase transitions in a
  near-critical solvent: A simple model reveals a rich phase diagram.
\newblock {\em Phys. Rev. Lett.}, 114:038301, Jan 2015.

\bibitem{BrunnerDobnikar04}
Matthias Brunner, Jure Dobnikar, Hans-Hennig von Gr\"unberg, and Clemens
  Bechinger.
\newblock Direct measurement of three-body interactions amongst charged
  colloids.
\newblock {\em Phys. Rev. Lett.}, 92:078301, Feb 2004.

\bibitem{SoykaZvyaHertHeldBech08}
Florian Soyka, Olga Zvyagolskaya, Christopher Hertlein, Laurent Helden, and
  Clemens Bechinger.
\newblock Critical {Casimir} forces in colloidal suspensions on chemically
  patterned surfaces.
\newblock {\em Phys. Rev. Lett.}, 101:208301, Nov 2008.

\bibitem{BonnOtwiSacaGuoWegSchall09}
Daniel Bonn, Jakub Otwinowski, Stefano Sacanna, Hua Guo, Gerard Wegdam, and
  Peter Schall.
\newblock Direct observation of colloidal aggregation by critical {Casimir}
  forces.
\newblock {\em Phys. Rev. Lett.}, 103:156101, Oct 2009.

\bibitem{ZAB11}
O.~Zvyagolskaya, A.~J. Archer, and C.~Bechinger.
\newblock Criticality and phase separation in a two-dimensional binary
  colloidal fluid induced by the solvent critical behavior.
\newblock {\em EPL (Europhysics Letters)}, 96(2):28005, 2011.

\bibitem{GZS12}
Nicoletta Gnan, Emanuela Zaccarelli, and Francesco Sciortino.
\newblock Tuning effective interactions close to the critical point in
  colloidal suspensions.
\newblock {\em The Journal of Chemical Physics}, 137(8):084903, 2012.

\bibitem{GZTS12}
Nicoletta Gnan, Emanuela Zaccarelli, Piero Tartaglia, and Francesco Sciortino.
\newblock How properties of interacting depletant particles control aggregation
  of hard-sphere colloids.
\newblock {\em Soft Matter}, 8:1991--1996, 2012.

\bibitem{DVNBS13}
Minh~Triet Dang, Ana~Vila Verde, Van~Duc Nguyen, Peter~G. Bolhuis, and Peter
  Schall.
\newblock Temperature-sensitive colloidal phase behavior induced by critical
  {Casimir} forces.
\newblock {\em The Journal of Chemical Physics}, 139(9):094903, 2013.

\bibitem{NguyenFaHuWeSch13}
Van~Duc Nguyen, Suzanne Faber, Zhibing Hu, Gerard~H. Wegdam, and Peter Schall.
\newblock Controlling colloidal phase transitions with critical {Casimir}
  forces.
\newblock {\em Nat. Commun.}, 4:1584, Mar 2013.

\bibitem{TasiosDijkstra17}
Nikos Tasios and Marjolein Dijkstra.
\newblock From {2D} to {3D}: Critical {Casimir} interactions and phase behavior
  of colloidal hard spheres in a near-critical solvent.
\newblock {\em The Journal of Chemical Physics}, 146(13):134903, 2017.

\bibitem{HobrechtHucht16a}
Hendrik Hobrecht and Alfred Hucht.
\newblock Critical {Casimir} force scaling functions of the two-dimensional
  {Ising} model at finite aspect ratios.
\newblock {\em J.~Stat. Mech.: Theory Exp.}, 2017:024002, Feb 2017.
\newblock {arXiv:1611.05622}.

\bibitem{HobrechtHucht18a}
Hendrik Hobrecht and Alfred Hucht.
\newblock Anisotropic scaling of the two-dimensional {Ising} model {I}: the
  torus.
\newblock {\em SciPost Phys.}, 7:26, Aug 2019.

\bibitem{HobrechtHucht18b}
Hendrik Hobrecht and Alfred Hucht.
\newblock Anisotropic scaling of the two-dimensional {Ising} model {II}:
  surfaces and boundary fields.
\newblock {\em SciPost Phys.}, 8:32, Mar 2020.

\bibitem{Lawden89}
Derek~F. Lawden.
\newblock {\em Elliptic Functions and Applications}.
\newblock Applied Mathematical Sciences. Springer-Verlag New York, 1989.

\bibitem{NIST}
Frank Olver, Daniel Lozier, Ronald Boisvert, and Charles Clark.
\newblock {\em {NIST Handbook of Mathematical Functions}}.
\newblock Jan 2010.

\bibitem{NIST:DLMF}
{{NIST Digital Library of Mathematical Functions}}.
\newblock http://dlmf.nist.gov/, Release 1.1.1 of 2021-03-15.
\newblock F.~W.~J. Olver, A.~B. {Olde Daalhuis}, D.~W. Lozier, B.~I. Schneider,
  R.~F. Boisvert, C.~W. Clark, B.~R. Miller, B.~V. Saunders, H.~S. Cohl, and
  M.~A. McClain, eds.

\bibitem{IorgovLisovyy11}
N.~Iorgov and O.~Lisovyy.
\newblock Ising correlations and elliptic determinants.
\newblock {\em J.~Stat. Phys.}, 143(1):33, 2011.

\bibitem{BultheelBarel}
A.~Bultheel and M.~Van Barel.
\newblock {\em Linear Algebra, Rational Approximation and Orthogonal
  Polynomials}.
\newblock North Holland, 2011.

\bibitem{MMA12}
{Wolfram Research, Inc.}
\newblock {\em Mathematica V12.2}.
\newblock Champaign, Illinois, 2020.

\bibitem{Parlett80}
Beresford~N. Parlett.
\newblock {\em The Symmetric Eigenvalue Problem}.
\newblock SIAM Classics in applied mathematics, 1980.

\bibitem{HeinigRost}
Georg Heinig and Karla Rost.
\newblock {\em Bezoutians}.
\newblock Technische Universit\"{a}t Chemnitz, Fakult\"{a}t f\"{u}r Mathematik
  (Germany), 2000.

\bibitem{LutherRost04}
Uwe Luther and Karla Rost.
\newblock Matrix exponentials and inversion of confluent {Vandermonde}
  matrices.
\newblock {\em Electron. Trans. Numer. Anal.}, 18:91--100, 2004.

\bibitem{FisherHartwig68}
M.~E. Fisher and R.~E. Hartwig.
\newblock Toeplitz determinants: some applications, theorems and conjectures.
\newblock {\em Adv. Chem. Phys.}, 15:333--353, 1968.

\bibitem{HartwigFisher69}
R.~E. Hartwig and M.~E. Fisher.
\newblock Asymptotic behavior of {Toeplitz} matrices and determinants.
\newblock {\em Arch. Rat. Mech. Anal.}, 32:190--225, 1969.

\bibitem{BasorEhrhardt02}
Estelle~L. Basor and Torsten Ehrhardt.
\newblock Some identities for determinants of structured matrices.
\newblock {\em Linear Algebra and its Applications}, 343-344:5--19, 2002.
\newblock Special Issue on Structured and Infinite Systems of Linear equations.

\bibitem{DeiftItsKrasovsky13}
Percy Deift, Alexander Its, and Igor Krasovsky.
\newblock Toeplitz matrices and {Toeplitz} determinants under the impetus of
  the {Ising} model: Some history and some recent results.
\newblock {\em Communications on Pure and Applied Mathematics},
  66(9):1360--1438, 2013.

\bibitem{Szego15}
G.~Szeg\H{o}.
\newblock {Ein Grenzwertsatz \"{u}ber die Toeplitzschen Determinanten einer
  reellen positiven Funktion}.
\newblock {\em Math. Ann.}, 76:490--503, 1915.

\bibitem{Szego20}
G.~Szeg\H{o}.
\newblock {Beitr\"{a}ge zur Theorie der Toeplitzschen Formen, I}.
\newblock {\em Math. Zeit.}, 6:167--202, 1920.

\end{thebibliography}

\end{document}